\documentstyle{mn}

\newif\ifAMStwofonts

\ifoldfss
  \ifCUPmtlplainloaded \else
    \NewTextAlphabet{textbfit} {cmbxti10} {}
    \NewTextAlphabet{textbfss} {cmssbx10} {}
    \NewMathAlphabet{mathbfit} {cmbxti10} {} 
    \NewMathAlphabet{mathbfss} {cmssbx10} {} 
  \fi
  \ifAMStwofonts
    \ifCUPmtlplainloaded \else
      \NewSymbolFont{upmath} {eurm10}
      \NewSymbolFont{AMSa} {msam10}
      \NewMathSymbol{\upi}     {0}{upmath}{19}
      \NewMathSymbol{\umu}     {0}{upmath}{16}
      \NewMathSymbol{\upartial}{0}{upmath}{40}
      \NewMathSymbol{\leqslant}{3}{AMSa}{36}
      \NewMathSymbol{\geqslant}{3}{AMSa}{3E}

    \fi
  \fi
\fi 

\ifnfssone
  \newmathalphabet{\mathit}
  \addtoversion{normal}{\mathit}{cmr}{m}{it}
  \addtoversion{bold}{\mathit}{cmr}{bx}{it}
  \newmathalphabet{\mathbfit} 
  \addtoversion{normal}{\mathbfit}{cmr}{bx}{it}
  \addtoversion{bold}{\mathbfit}{cmr}{bx}{it}
  \newmathalphabet{\mathbfss} 
  \addtoversion{normal}{\mathbfss}{cmss}{bx}{n}
  \addtoversion{bold}{\mathbfss}{cmss}{bx}{n}
  \ifAMStwofonts
    \ifCUPmtlplainloaded \else
      \UseAMStwoboldmath
      \makeatletter
      \new@mathgroup\upmath@group
      \define@mathgroup\mv@normal\upmath@group{eur}{m}{n}
      \define@mathgroup\mv@bold\upmath@group{eur}{b}{n}
      \edef\UPM{\hexnumber\upmath@group}
      \new@mathgroup\amsa@group
      \define@mathgroup\mv@normal\amsa@group{msa}{m}{n}
      \define@mathgroup\mv@bold\amsa@group{msa}{m}{n}
      \edef\AMSa{\hexnumber\amsa@group}
      \makeatother
      \mathchardef\upi="0\UPM19
      \mathchardef\umu="0\UPM16
      \mathchardef\upartial="0\UPM40
      \mathchardef\leqslant="3\AMSa36
      \mathchardef\geqslant="3\AMSa3E
    \fi
  \fi
\fi 

\ifnfsstwo
  \DeclareMathAlphabet{\mathbfit}{OT1}{cmr}{bx}{it}
  \SetMathAlphabet\mathbfit{bold}{OT1}{cmr}{bx}{it}
  \DeclareMathAlphabet{\mathbfss}{OT1}{cmss}{bx}{n}
  \SetMathAlphabet\mathbfss{bold}{OT1}{cmss}{bx}{n}
  \ifAMStwofonts
    \ifCUPmtlplainloaded \else
      \DeclareSymbolFont{UPM}{U}{eur}{m}{n}
      \SetSymbolFont{UPM}{bold}{U}{eur}{b}{n}
      \DeclareSymbolFont{AMSa}{U}{msa}{m}{n}
      \DeclareMathSymbol{\upi}{0}{UPM}{"19}
      \DeclareMathSymbol{\umu}{0}{UPM}{"16}
      \DeclareMathSymbol{\upartial}{0}{UPM}{"40}
      \DeclareMathSymbol{\leqslant}{3}{AMSa}{"36}
      \DeclareMathSymbol{\geqslant}{3}{AMSa}{"3E}
    \fi
  \fi
\fi 

\ifCUPmtlplainloaded \else
  \ifAMStwofonts \else 
    \def\upi{\pi}
    \def\umu{\mu}
    \def\upartial{\partial}
  \fi
\fi

\title[SwSt~1: the star and the nebula]{SwSt~1: an O-rich planetary nebula \hfill \break around a C-rich central star} 
\author [O. De Marco et al.]{Orsola De Marco$^{1,2,3}$, P.A. Crowther$^3$, M.J. Barlow$^3$, \cr
        Geoffrey C. Clayton$^2$, A.~de~Koter$^4$\\
$^1$Dept. of Astrophysics, American Museum of Natural History, Central Park West at 79$^{\rm th}$ Str., New York, NY 10024, USA \\
$^2$Dept. of Physics and Astronomy, Louisiana State University, Baton Rouge, LA 70803, USA \\
$^3$Dept. of Physics and Astronomy, University College London, Gower Str., London WC1E 6BT, UK \\
$^4$Astronomical Institute `Anton Pannekoek', University of Amsterdam, Kruislaan 403, 1098 Amsterdam, The Netherlands}

\date{Received February 2001; }

\pagerange{\pageref{firstpage}--\pageref{lastpage}}
\pubyear{1999}

\begin{document}

\maketitle

\label{firstpage}

\begin{abstract}
The hydrogen-deficient [WCL] type
central star HD167362 and its planetary nebula (PN)
SwSt~1 are investigated. 
The central star has a carbon-rich emission
line spectrum and yet the nebula exhibits a 10-$\mu$m emission feature
from warm silicate dust, perhaps indicating a recent origin for the
carbon-rich stellar spectrum.
Its stellar and nebular properties might therefore
provide further understanding as to the origin of the [WCL] central star class.

The central star optical and UV spectra are modelled
with state of the art non-LTE codes for expanding atmospheres, from which
the stellar parameters are determined. Using the Sobolev approximation code {\sc isa}-Wind,
we find T$_{\rm eff}$=40\,000~K,
log(${\dot M}$/M$_\odot$~yr$^{-1}$)=--6.72, L=8900~L$_\odot$ (for a distance
of 2.0~kpc), and $v_\infty$ $\simeq$ 900~km~s$^{-1}$. The abundance mass fractions 
for helium, carbon and oxygen is determined to be 37\%, 51\% and 12\%, respectively.
From this we derive C/O=4.3 (by mass), confirming that the star suffered efficient 
third dredge-up.
The nitrogen abundance is close to zero, while an upper limit of $<$10\% by mass is 
established for H.
The model
uses a composite beta velocity law which allows us to reproduce the optical line profiles.
The overall shape of the de-reddened spectrum agrees with the 
V-scaled ($m_V$=11.48~mag, E(B--V)=0.46~mag) model atmosphere, showing the
nebular-derived reddening to be consistent with the reddening indicated 
by the stellar analysis. We confirm our model results by using the co-moving frame code
{\sc cmfgen}, although a few differences remain.

The PN has a high electron density (log(N$_e$/cm$^{-3}$)=4.5) and a small
ionized radius (0.65~arcsec - measured from the HST-WF/PC H$\beta$
images), indicating a young object. Its nebular abundances are
not peculiar. The nebular C/O ratio is close to solar,
confirming the PN as an O-rich nebula. The nebular N/O ratio of
0.08 is {\it not} indicative
of a Type-I PN although the high stellar luminosity points to a relatively
stellar mass.
Near-IR spectroscopy is presented and fitted together with IRAS fluxes
by using two blackbody curves with temperatures 1200~K and 230~K, indicating the
presence of hot dust. We also report the first detection of H$_2$ in this
young and compact PN.

All of the published spectroscopy since the discovery of SwSt~1 in 1895
has been re-examined and it is concluded that no clear spectral
variability is seen, in contrast to claims
in some previously-published studies. If an event occurred that
has turned it into a 
hydrogen-deficient central star, it did not happen in the last 100 years.

\end{abstract}


\section{Introduction}
\label{sec:introduction}

[WCL]\footnote{ `WC' stands for Wolf-Rayet of the carbon sequence, `L'
stands for `late' as opposed to the hotter members of this class, the
[WCE]. The brackets were introduced by van der Hucht et al. (1981) to
differentiate Wolf-Rayet central stars from massive central stars.}
central stars of Planetary Nebulae (PN) are hydrogen-deficient central
stars that exhibit emission line spectra very similar to massive
Wolf-Rayet stars.  Until recently it has been impossible to calculate
stellar evolutionary models where the star eliminates all of its
hydrogen-rich envelop at the top of the Asymptotic Giant Branch (AGB).
This prompted the association of these objects with the born-again
scenario
(Iben et al. 1983), whereby a white dwarf experiences a last helium shell
flash, after which it rejoins the tip of the AGB for a second time and
repeats its evolution, this time as a hydrogen-deficient central star.

The PN around [WC] central stars do not appear to have any characteristics
which distinguish them from the PN of hydrogen-rich central stars. A
marginally higher nebular C/H ratio (De Marco \& Barlow 2001), as well as
a slightly higher PN expansion velocity (Gorny \& Stasinska 1995) , do not
appear to be sufficient to characterise the sample's evolution. Other
puzzling peculiarities concern spectral appearance differences amongst
members of the hydrogen-deficient central star group. Some hydrogen
deficient central stars appear to have weaker lines than others (called
{\it weak emission lines stars} ({\it wels}) by Tylenda, Acker \& Stenholm
(1993)), while in the strong-lined [WC] group, there appears to be a
classification gap, with [WC] stars populating the [WC11-8] class (called
[WCL]) and [WC5-WO1] (called [WCE]; Pottasch 1996, Crowther, De Marco \&
Barlow 1998). Additionally, while it is currently believed that [WCL]
stars
evolve into [WCE] stars, the stellar abundances seem to differ in the two
groups, with the more evolved group having a lower carbon abundance than
the less evolved group (see De Marco \& Barlow (2001) for a summary of
results from the literature), a fact that cannot be easily explained.

On the other hand, Infrared Space Observatory (ISO) spectroscopy of PN
with [WCL] central stars has shown that all have crystalline silicate
features in their mid- to far-infrared spectra, in addition to the
previously known carbon-rich PAH-type features that dominate their
mid-infrared spectra, betraying the simultaneous presence of both
carbon-rich and oxygen-rich dust (see Barlow 1997, Waters et al. 1998,
Cohen et al. 1999, Cohen 2001). The relatively warm temperatures deduced
for the O-rich dust, together with the high nebular densities of their
compact nebulae, appear to imply that the C-rich central stars evolved
relatively recently from an O-rich composition. This would in turn suggest
that these [WCL] PN have been on the AGB recently and so did not result
from a born-again evolutionary phase, for which low-density nebulae are
expected. This would also imply that the carbon-rich [WCL] stars systematically
result from oxygen-rich AGB giants.

Recent improvements in theoretical stellar evolutionary calculations have
succeeded in eliminating hydrogen from the stellar atmosphere at the tip
of the AGB, so that the requirement that the central star undergo a
born-again event in order to get rid of its hydrogen is no longer in
place. According to Herwig (2000) a thermal pulse just before the star
departs from the AGB can eliminate most of the hydrogen, while a
well-timed post-AGB pulse (probabilistically rarer) could account for the
extreme hydrogen deficiency observed in [WCL] stars. The combination of
the ISO observations and the new evolutionary calculations therefore
contributes to the revision of theories about the origin of [WCL]
central stars and for the first time we may have a viable scenario that
does not require the star to go through a born-again episode.

The PN SwSt~1 around the central star HD~167362 (we will
give both star
and nebula the name SwSt~1) exhibits a broad 10$\mu$m silicate emission
feature (Aitken \& Roche 1982) and has been suggested to be amongst the
youngest of 
[WCL] nebulae, possibly having experienced the transition from the AGB in
the last
century.  The spectrum of SwSt~1 was first reported by Fleming (1895),
but the first
complete spectroscopic analysis was carried out by Swings and Struve
(1940, 1943)\nocite{SS40}\nocite{SS43}.  The traditional WR 
C~{\sc iv} $\lambda$5806/C~{\sc iii} $\lambda$5696 line diagnostic ratio 
positions it within the [WC9] spectral class (Crowther et al. 1998 - see
also previous classifications by Cohen (1975\nocite{C75}) and by Carlson
and Henize (1979\nocite{CH79})), but the weakness of its lines compared to
other [WC9] stars (e.g. BD+303639, He~2-99) or even cooler [WC10] stars
(e.g. CPD--56$^{\rm o}$8032, He~2-113, M~4-18) prompted Crowther et
al. (1998) to flag it as {\it peculiar} and has therefore essentially left
it in a class of its own.  Tylenda et al. (1993) listed it amongst
their {\it wels} because of the weakness of its spectral lines, but
SwSt~1 does not resemble stars in this class either (e.g. Cn~3-1,
De Marco 2000) since it has many more stellar emission
features than any {\it wels}, indicative of a denser wind.


In this paper we tried to determine whether the physical characteristics
of the central star and the PN can shed light on the reason for these
differences, i.e. whether SwSt~1 is at an intermediate phase between stars
in the [WCL] class and their ancestors.
In the Sections that follow we present optical echelle spectra,
near-infrared spectra and archival IUE spectra, as well as HST images of
this central star and its PN. Using these data we carry out a
thorough analysis, to try and single out the peculiarities of this object.  
In Section~2 we present our observational data, while in Section~3 we
derive some observational parameters. The distance, luminosity and
mass are discussed in Section~4. The stellar analysis is tackled in
Section~5, while the nebular analysis is carried out in Sections~6 to 8.
Finally (Section~9) a review of the secular variability is reported and
conclusions are drawn in Section~10.

\section{Observations}
\label{sec:observations}

SwSt~1 was observed with the 3.9~m
Anglo-Australian Telescope (AAT) on May 14 and 15,
1993, using the 31.6 l/mm grating of the UCL Echelle Spectrograph (UCLES), 
with a 1024$\times$1024 pixel Tektronix
CCD as detector. Four settings of the CCD were needed in order to obtain
spectral
coverage in the range 3600--9700~\AA .
At each wavelength setting, 5~arcsec wide-slit spectra were obtained
for absolute
spectro-photometry, while 1.5~arcsec narrow--slit spectra 
(R=50,000) were obtained for maximum resolution in the range
4100--5400~\AA .
The continuum signal-to-noise ratio of the averaged exposures
ranges from $\sim$5 in the far blue to $\sim$30 in the red.
A log
of the observations is presented in Table~\ref{tab:aat93_log}.     

The data were reduced using the {\sc iraf} (V2.10)\footnote{
IRAF is written and supported by the National Optical Astronomy Observatories
(NOAO) in Tucson, Arizona; http://iraf.noao.edu/ }
package at the UCL Starlink node. Wavelength
calibration was with respect to comparison Th-Ar arc exposures. Absolute
flux calibration 
was with respect to the B3{\sc iv} star HD~60753, whose energy
distribution has been measured by Oke (1992)\nocite{O74}\nocite{O92}.
Details of the laborious echelle data flux calibration procedure can be
found in De Marco, Barlow \& Storey (1997; hereafter DBS97).
Subsequent data reduction
was carried out with the Starlink {\sc dipso} package (Howarth et al.     
1995).
After flux calibration, the narrow--slit spectra were scaled up to
the continuum level of the wide--slit spectra.

\begin{table}
\caption{Log of the 1993 UCLES observations of SwSt~1.  
Only the second night was photometric.}
\begin{center}
\begin{tabular}{cccccc}
Central & Date &  Exp. & Slit & Airmass & Seeing \\
$\lambda$ &1993,           &  Time    & Width&         &       \\
  (\AA)   & May     &           (sec)   &(arcsec)     &    & (arcsec) \\
4622  & 14 & 2$\times$900 & 1.5 & 1.2 & 3\\
4622  & 14 &  300 & 5.0 & 1.14 & 2.5\\
6827  & 15 &  900 & 5.0 & 1.6 & 2 \\
6960  & 15 &  600 & 5.0 & 1.5 & 2 \\
4622  & 15 &  300 & 5.0 & 1.4 & 2 \\
3712  & 15 &  500 & 5.0 & 1.4 & 2 \\
6827  & 15 &   60 & 5.0 & 1.3 & 2 \\
6960  & 15 &   60 & 5.0 & 1.3 & 2 \\
\label{tab:aat93_log}
\end{tabular}
\end{center}
\end{table}      

International Ultraviolet Explorer (IUE) observations of SwSt~1 were also used.
The retrieved LWR and SWP spectra were divided up according to the
aperture used for
their acquisition and 
the weighted average used.
One SWP observation (SWP17068) was taken in high resolution mode and
although
its signal-to-noise ratio was rather poor, it could be used for parts of
our analysis.


\subsection{Near-IR Spectroscopy: First Detection of H$_2$ in a Young 
[WCL]-Type Central Star}\label{ssec:near_ir_spectroscopy}

Long slit, near-IR spectroscopy of SwSt~1 was acquired on Sept 1, 1999
with
the NTT, using the Son OF Isaac (SOFI) instrument, a 1024$\times$1024 pixel
NICMOS detector, and low resolution IJ (GRB) and HK (GRR)
gratings. The spectral coverage was 0.94--1.65~$\mu$m and
1.50--2.54~$\mu$m, respectively, with dispersions of 7.0~\AA/pix and
10.2~\AA/pix. The 0.6~arcsec slit provided a 2-pixel spectral resolution
of 14--20~\AA. The total integration time was 480 sec at each grating
setting. The removal of atmospheric features was achieved by observing
HD~169101 (A0V)
immediately before and after SwSt~1, at a close airmass (within 0.03).
Similar observations of HD\,166733 (F8) permitted a relative flux
correction, using a $T$=6100~K model atmosphere normalized to V=9.59~mag.

A standard extraction and wavelength calibration was carried out with
{\sc iraf}, while {\sc figaro} (Shortridge, Meyerdierks \& Currie, 1999)
and {\sc dipso} (Howarth et al. 1995) were used for the  atmospheric
and flux calibration, first artificially removing stellar hydrogen features
from the B3V spectrum. Our relatively fluxed
dataset was adjusted to match previously published near-IR photometry
(Allen \& Glass 1974) via convolution with the appropriate filter
profiles. The spectrum is presented in
Appendix~\ref{app:the_optical_and_near_ir_spectra_of_swst1}.

We can also report on the first detection of H$_2$, at 2.1218~$\mu$m
($v$=1--0, S(1)) from this young, late WC-type central star PN
system. Although the presence of $H_{2}$ is
well established in slightly more evolved [WC] systems such as the
[WC9] PN BD+30$^{\rm o}$3639 (Kastner
et al. 1996), it has not been detected from some young,
high-density PN with bright central stars, e.g. IC 4997 and IC 418
(Zuckerman \& Gatley 1988, Kastner et al. 1996). This appears to be due to
their strong $H_{2}$-dissociating UV radiation fields. 
The relative weakness of SwSt~1's H$_2$ emission may indicate that
the neutral regions
surrounding this nebula are also mainly atomic rather than molecular.
This is in agreement with the fact that no CO is detected in the neutral envelope of SwSt~1
(Huggins et al. 1996).

\section{Observational quantities}
\label{sec:observational_quantities}

\subsection{Apparent Magnitudes}
\label{ssec:apparent_magnitudes}

We convolved our observed central star wide slit spectro-photometry
of SwSt~1
with broad--band V and B filter profiles (kindly made available to
us by Dr. J.R. Deacon) to obtain
an estimate of the apparent V and B magnitudes of the stars.
We derive V=11.48~mag and B=11.54~mag. Convolving with narrow-band (Smith 1968)
filter profiles, we derive $v$=11.69~mag and $b$=11.73~mag.
The {\sc hipparcos} and {\sc tycho} catalogues quote a Johnson V magnitude
of 10.94~mag, Hpmag = 10.86~mag (median magnitude
in {\sc hipparcos} system) and an observed Hp range of 10.79--10.97~mag.
However, SwSt~1 is near the faint limit of the {\sc tycho} detector
of $\sim$12~ mag. 
The visual photometry of A. Jones, reveals a mean magnitude of 
(10.4 $\pm$ 0.5)~mag with no sign of variability (Jones et al. 1999).

We finally decided to adopt our own spectro-photometric values since
our spectrophotometry aligns perfectly with IUE
{\it and} near-IR SOFI data and since broadband photometry of emission
line central
stars can be contaminated by stellar and nebular emission line fluxes. 
Although the visual photometry of Jones et al. (1999) shows the
lack of intrinsic stellar variability, photometry of SwSt~1
carried out by A. Landolt (priv. comm.) on February 19, 20 and 23, 2001, indicates
V=11.57~mag (in the standard system of Landolt 1992)
\footnote{This is the average of six estimates, two per night in the range 11.55--11.59~mag. 
The aperture used was 14~arcsec wide. Two or three faint stars
were included in the aperture, such that this estimate might be slightly
too high, yet it is fainter than anything reported before. From the same
photometry the following colours are reported:  B--V = +0.078, U--B =
--0.960, V--R = +1.283, and R--I = --0.498.}, slightly fainter than our own
spectrophotometry as well as all the other estimates and indicating that
variability might be present.


\subsection{Nebular Radial and Expansion Velocities}
\label{ssec:nebular_radial_and_expansion_velocities}
\begin{table}
\caption
{Radial and expansion velocities (Columns 1 and 2) for SwSt~1
obtained from the Balmer emission lines.
Radial velocities for the Na~{\sc i} D absorption line components are
also presented (Column~3).
Finally, distances derived from two different methods are presented in
Columns~4 and 5.}
\renewcommand\arraystretch{0.7}
\begin{tabular}{lcccc}
LSR RV        & PN Expansion & LSR RV of   & D$^a$    & D$^b$  \\
              & Velocity     & Na~{\sc i}  &          &        \\
(km~s$^{-1}$) & (km~s$^{-1}$)&(km~s$^{-1}$)& (kpc)    & (kpc)  \\
--8.1 $\pm$ 1.0 & 21 $\pm$ 2     &--78,--38,+8 & 4.9 & 3.0   \\
\noalign{\smallskip}
\multicolumn{5}{l}{$^a$Distance derived from Na~{\sc i} D positive line components}\\
\multicolumn{5}{l}{$^b$Distance derived from the IR flux method}\\
\end{tabular}
\label{tab:radial_velocities}
\end{table}

Nebular radial velocities were obtained by determining
the average wavelength shift of the nebular Balmer lines using
a single Gaussian fit. Even when
a single Gaussian did not fit the data very well, 
it was felt that this would be the best way to
locate the overall centre of the line without biases introduced 
by asymmetry.
The heliocentric radial velocity was measured to be 
(--17.8 $\pm$ 1.0)~km~s$^{-1}$
from an average of the $H\beta$, H$\gamma$ and H$\delta$ lines;
the LSR radial velocity was determined to be (--8.1 $\pm$ 1.0)~km~s$^{-1}$.
These estimates compare well with those of Schneider et al. (1983;
$V_{\rm LSR}$=--8.7~km~s$^{-1}$, $V_{\rm Helio}$=--18.6~km~s$^{-1}$).
Radial velocities are listed in Table~\ref{tab:radial_velocities},
together with nebular expansion velocities measured from averaging the 
HWHM of $H\beta$, H$\gamma$ and H$\delta$ (H$\alpha$ was saturated).

\subsection{Reddening}
\label{ssec:reddening}

\begin{table}
\caption
{Wide--slit (5~arcsec) fluxes (ergs cm$^{-2}$ s$^{-1}$) measured for the nebular hydrogen lines of SwSt~1,
compared with values from the literature.}
\renewcommand\arraystretch{0.7}
\begin{small}
\begin{center}
\begin{tabular}{@{}lccc@{}}
Reference & H$\beta$  &H$\gamma$ &H$\delta$ \\
AAT (1993)&  3.19$\times$10$^{-11}$& 1.20$\times$10$^{-11}$& 6.53$\times$10$^{-12}$\\
Perek (1971) & 1.66$\times$10$^{-11}$& -- & -- \\
FGC84 &  2.14$\times$10$^{-11}$& 7.24$\times$10$^{-12}$& --           \\
dFV87 & 1.82$\times$10$^{-11}$ & -- & -- \\
AST89 & 5.89$\times$10$^{-11}$ & -- & -- \\
\noalign{\smallskip}
\multicolumn{4}{l}{FGC84 = Flower, Goharji and Cohen (1984)}\\
\multicolumn{4}{l}{dFV87 = de Freitas Pacheco and Veliz (1987)}\\
\multicolumn{4}{l}{AST89 = Acker, Stenholm, Tylenda (1989)}\\
\end{tabular}
\label{tab:hbeta_fluxes}
\end{center}
\end{small}
\end{table}

\begin{table}
\caption{Reddenings derived from Balmer lines ratios
and from the radio--H$\beta$ baseline for SwSt~1.}
\renewcommand\arraystretch{0.7}
\begin{tabular}{ccccc}
Ratio & E(B--V)  & E(B--V)  & E(B--V) & E(B--V) \\
                         &AAT (1993) & FGC84 & dFP87 & P71 \\
H$\alpha$/H$\beta$& --       &  --    & -- & -- \\
H$\beta$/H$\gamma$& 0.51$^a$ &  --    & -- & -- \\
H$\beta$/H$\delta$& 0.41$^a$ &  --    & -- & -- \\
radio/H$\beta$    & 0.30     &  0.34  & 0.41 & 0.50$^b$ \\
\noalign{\smallskip}
\multicolumn{5}{l}{FGC84 = Flower, Goharji and Cohen 1984}\\
\multicolumn{5}{l}{GD85 = Goodrich and Dahari 1985}\\
\multicolumn{5}{l}{dFV87 = de Freitas Pacheco and Veliz 1987}\\
\multicolumn{5}{l}{P71 = Perek 1971}\\
\multicolumn{5}{l}{$^a$ We adopted 0.46, the average of 0.51 and 0.41}\\
\multicolumn{5}{l}{$^b$ Derived by us using his H$\beta$ flux}\\
\end{tabular}
\label{tab:reddening}
\end{table}

The reddening was
derived by comparing the relative fluxes in the nebular 
H$\beta$, H$\gamma$ and H$\delta$ lines (H$\alpha$ was saturated).
Using the Case B hydrogen recombination
coefficients of Storey and Hummer (1995)\nocite{SH95} for a nebula with
electron density N$_e$ = 10$^4$
cm$^{-3}$ and T$_e$ = 10$^4$ K and the Galactic reddening law
of Howarth (1983)\nocite{H83}, values of E(B--V) were derived.
The reddening was also derived from a comparison of the radio and
H$\beta$ fluxes, using the method of Milne and Aller (1975)\nocite{MA75}.
In Table~\ref{tab:hbeta_fluxes} we list Balmer line fluxes measured from
wide-slit
AAT spectroscopy, together with fluxes from the literature.
Considerable discrepancies exist between the H$\beta$ flux measured here
and measurements found in the literature.
From Table~\ref{tab:hbeta_fluxes} we see that our own value is about one third
higher than most of the other values, although Acker, Stenholm and Tylenda
(1989)\nocite{AST89}
measured an even higher H$\beta$ flux. The possibility of a 
true variation in time was therefore investigated, 
but no clear trend exists in either the H$\beta$ fluxes or in the radio
flux measurements. Although these large differences remain suspicious, 
we must attribute them to calibration and measurement errors
and possibly to some slit/aperture widths not including all of the PN.

In Table~\ref{tab:reddening} we list our results together with reddenings
selected from the literature for comparison. From the H$\beta$--H$\gamma$
baseline in our UCLES wide--slit (5~arcsec) spectrum, we determine
E(B--V)=0.51, while from the H$\beta$--H$\delta$ baseline we find
E(B--V)=0.41. Radio fluxes at 15~GHz have been measured by Kwok et al.
(1981; (207 $\pm$ 11)~mJy), Milne and Aller \nocite{MA82} (1982; (240 $\pm$
12)~mJy) and Aaquist and Kwok (1990; (175 $\pm$ 17)~mJy). The
weighted average ((211 $\pm\ 17$)~mJy) corresponds to a 5~GHz flux of 243.0
mJy for optically thin free--free emission, which we adopted, together
with our UCLES value for the H$\beta$ flux, to derive E(B--V)=0.30. A
similar value is also obtained by nullying the 2200~\AA\ feature.
Comparison of the Perek (1971)\nocite{P71} photo-electric flux with the
mean radio flux yields E(B--V)=0.50.

We adopt E(B--V)=0.46$\pm$0.05,
the mean obtained from the H$\beta$/H$\gamma$
and H$\beta$/H$\delta$ ratios, as the most self-consistent estimate we have. 
Pre-emptying the results of Sec.~\ref{ssec:stellar_model_results}, the
slope of the synthetic stellar atmosphere model carried out with the {\sc isa}-Wind code,
which fits the stellar
emission line intensities, agrees with the nebular line reddening estimate
of E(B--V)=0.46. For the model carried out with the {\sc cmfgen} code, 
the best match is obtained with 
E(B--V)=0.48. 
These values are also in reasonable agreement with the determination
of E(B--V)=0.50 quoted above, from the radio and the photo-electric
H$\beta$ fluxes.

Finally, the near-IR spectrum is dominated
by strong nebular emission lines of H\,{\sc i} (e.g. P$\beta$, Br$\gamma$) 
and can be used to obtain a further confirmation of the reddening.
For E(B--V)=0.46 and the nebular
properties derived in Sec.~\ref{sec:nebular_abundance_analysis}, 
agreement with Case~B recombination theory (Storey \& Hummer
1995) is very good. For instance,
$I$(P$\beta)/I$(H$\beta$)=0.16 (0.15 from Case B), or
$I$(Br$\gamma)/I$(H$\beta$)=0.024 (0.024  from Case B).
P$\alpha$ is an exception, but this lies in a region of poor
atmospheric transmission.

\section{Luminosity and Distance}
\label{sec:luminosity_and_distance}

\begin{table*}
\begin{minipage}{130mm}
\caption{Summary of the distances determined for SwSt~1 in the literature
using various methods.}
\begin{tabular}{lll}
Distance & Method & Reference \\
(kpc)    &        &      \\
1.0 & extinction method   & de~Freitas Pacheco \& Veliz 1987\\
1.1 & modified Shklowskii & de~Freitas Pacheco \& Veliz 1987\\
$>$1.2 & {\sc hipparcos}     & Acker et al. 1998\\
1.4 & Shklowskii          & Cahn et al. 1992 \\
1.4 & M$_v$ - T$_*$ relationship & de~Freitas Pacheco \& Veliz 1987\\
3.0 & LMC [WC] star calibration& this work \\
3.1 & M and gravity           &  Zhang 1993 \\  
3.2 & I(H$\beta$) for opt. thick PN$^a$ & Kingsburgh \& Barlow 1992\\
3.8 & ionized mass - R(PN) and radio Tb - R(PN) corr & Zhang 1995 \\
4.7 & radio cont. brightness and R(PN) & van de Steene and Zijlstra 1994 \\
4.9 & Galactic rotation curve & this work \\
5.0 & L$_*$               &  Zhang 1993 \\  
\noalign{\smallskip}
\multicolumn{3}{l}{$^a$We used their Eq.~11 and our 
log(I(H$\beta$))=--10.1 from Sec.~\ref{ssec:zanstra_temperatures}}
\end{tabular}
\label{tab:distances}
\end{minipage}
\end{table*}

In Table~\ref{tab:distances} we summarise the distances that have been
determined for SwSt~1 in the literature, together with our two estimates
from
Sections~\ref{ssec:distance_from_a_magellanic_cloud_luminosity_calibration} 
and \ref{ssec:galactic_rotation_curve_distance}. Below we explain our two
methods but admit that there is no convincing estimate for the
distance to SwSt~1 beside the fact that after iterating between an
assumed distance
and our models 
we have gained further insight which restricts the range of possible
distances (see Secs.~\ref{sec:stellar_analysis} and
\ref{sec:nebular_abundance_analysis}), leading us to finally adopt a
distance of 2.0~kpc (Section~\ref{ssec:adopted_distance}).

\subsection{Distance from a Magellanic Cloud Luminosity Calibration}
\label{ssec:distance_from_a_magellanic_cloud_luminosity_calibration}

DBS97 used Magellanic Cloud Wolf--Rayet
central stars to derive a mean central star mass for WC central stars 
of 0.62~M$_\odot$.
This mass was then applied together with the helium-burning tracks of
Vassiliadis \& Wood (1994) to derive luminosities for 
CPD--56$^{\rm o}$8032 and He~2--113
corresponding to their effective temperatures, which were in turn used,
together with their integrated IR fluxes, to determine their distances.
The PN of SwSt~1, with a high electron density, a small apparent radius 
and a significant IR
excess appears to be a good candidate to apply the same method. 
We have obtained from the ISO archive the SWS01 spectrum of SwSt~1
(from the programme of S.K. Gorny, TDT47101511, reduced with SWS OLP
Version 9.5, Szczerba et al. 2001). This spectrum is plotted in Fig.~\ref{fig:model_flux},
together with its photometric fluxes at 1.65 and 2.2~$\mu$m (Allen \& Glass 1974) 
and colour-corrected IRAS fluxes. 
Integration under the IR spectrum of SwSt~1 between 1.6 and 100~$\mu$m
results in a flux of 1.48$\times$ 10$^{-8}$ erg~s$^{-1}$~cm$^{-2}$,
corresponding 
to an infrared bolometric luminosity of 463~D$^2$(kpc)~L$_\odot$.
For T$_{\rm eff}$$\sim$40000~K (Section~\ref{sec:stellar_analysis}),
the helium--burning tracks of Vassiliadis and Wood (1994)\nocite{VW94}
predict a luminosity of 4280~L$_{\odot}$, where we interpolated between the
tracks for 0.600 M$_\odot$ and 0.634 M$_\odot$ cores. 
The distance obtained in this way is 3.0~kpc (a T$_{\rm eff}$$\sim$35000~K
predicts 4670 L$_{\odot}$ and a distance of 3.2~kpc).

The three key assumptions of this method are: (a) SwSt~1 has the same mass
as the LMC [WC] stars, (b) all its luminosity is re-radiated in the 
IR by dust and (c) the helium-burning track calculations are valid.
By far the most compromising assumption is that all [WC] stars have the
same mass
(as suggested by De Marco and Crowther (1998) this is likely not to be
the case).
We have therefore determined the distances corresponding to all three masses
calculated by Vassiliadis and Wood (1994), namely 0.57, 0.60 and 0.63~M$_\odot$.
These are 2.1, 2.5 and 3.6~kpc, respectively. We therefore conclude that
if
assumptions (b) and (c) hold, the distance is in the range
$\sim$2.0--3.5~kpc.

\subsection{Galactic Rotation Curve Distance}
\label{ssec:galactic_rotation_curve_distance}

\begin{figure*}
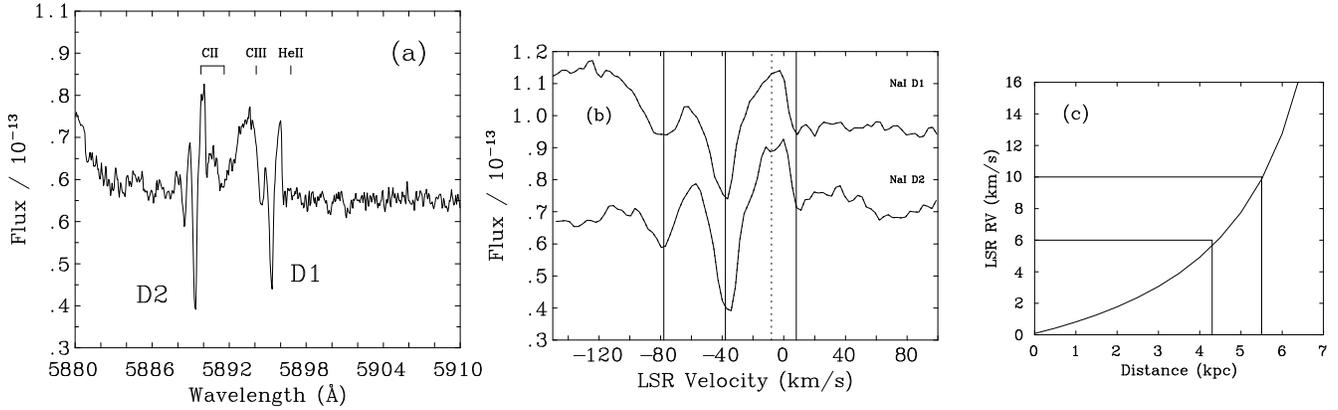

\vspace{5cm}
\caption
{(a) The Na~{\sc i} D lines towards SwSt~1 (where the wavelength scale has been corrected for
a heliocentric radial velocity of
--17.8 km~s$^{-1}$ and where we have indicated the positions of C~{\sc ii} Multiplet 5,
C~{\sc iii} Multiplet 20 and the He~{\sc ii} $\lambda$5896.8 line.
(b) The Na~{\sc i} D lines
displayed in their LSR rest frame. The D2 line has its original flux units,
while the D1 line has been displaced upwards
by an additive factor of 3$\times$10$^{-14}$. The dotted line marks the LSR radial velocity
of the nebula, while the solid lines show the position of the three most prominent Na~{\sc i}
absorption components.
(c) The Galactic rotation curve for the direction of SwSt~1. The distances corresponding to LSR radial velocities
of +6 and +10 km~s$^{-1}$ are 4.3 and 5.5 kpc. }
\includegraphics{naD_lines_spec.eps}
\includegraphics{naD_lines_vel.eps}
\includegraphics{naD_lines_rot.eps}
\label{fig:na_lines}
\end{figure*}

In order to obtain an alternative value for the distance,
we applied the insterstellar sodium D line method used by DBS97.
In Fig.~\ref{fig:na_lines}(a) we show the spectral range containing the Na~{\sc i} D lines,
while in Fig.~\ref{fig:na_lines}(b) we show Na~{\sc i} D1 and D2 lines in their respective LSR rest frames.
From Fig.~\ref{fig:na_lines}(b) and (c) we see that the
line of sight to SwSt~1 has a positive rotation curve, while most of 
the Na~{\sc i} D line components have negative radial shifts. 
The only interstellar component with positive
velocity is measured at (+8 $\pm$ 2)~km~s$^{-1}$ 
(see Fig.~\ref{fig:na_lines}(b)) and corresponds to a distance of (4.9 $\pm$ 0.6)~kpc,
where the uncertainty derives solely from locating the centre of the trough.
The underlying stellar emission lines (C~{\sc ii}
$\lambda$$\lambda$5889,92, C~{\sc iii} $\lambda$5894 and He~{\sc ii} $\lambda$5897)
are too weak to achieve a reliable rectification, which would lead to an increase in the
measurement precision. Moreover the Na~{\sc i}
D lines have emission components (as noticed by Dinerstein et al. (1995))
and a multiple cloud fit would over-interpret
the data, so we limited ourselves to measuring the centres of the
absorption components.
The negative radial velocity troughs are likely to derive from the PN's
neutral envelope
and are discussed in Appendix~\ref{app:the_neutral_envelope_of_swst1}.

\subsection{The Adopted Distance}
\label{ssec:adopted_distance}

Although each method listed in Table~\ref{tab:distances} has its limitations and it is beyond the scope of this
paper to review each of the different estimates, we should note that two 
of the best methods to estimate distances to Galactic objects
suffer from additional caveats. SwSt~1 is too far for the {\sc hipparcos}
distance, derived from a parallax of (8.9 $\pm$ 6.0)~mas (HIP89535; Acker et
al. 1998),
to be considered reliable.  
Second, the extinction distance is based on the
assumption that within 2~kpc of the sun, light suffers 0.4~mag of
extinction per kpc, and is not
based on an extinction map, which would make the method more reliable. 


Although our estimates seem to point to a fairly large distance (i.e.
$\sim$3--5~kpc), we found in our modelling of the stellar spectrum that
a very high luminosity and central star mass were implied by a
distance larger than 2~kpc and would have required the central star to
have evolved
significantly in effective temperature in the last 100 years. Such
evolution would have been
mirrored by significant spectral changes. This however has not been
observed (see Sec.~\ref{sec:swst1_between_1895_and_1993}).  We therefore
adopt 2~kpc as the distance to SwSt~1, but admit a substantial
uncertainty.


\section{Stellar Analysis}
\label{sec:stellar_analysis}

In this Section we carry out a stellar atmosphere analysis of SwSt~1 to
derive the stellar effective temperature and mass-loss rate as well
as its carbon and oxygen abundances. The line-blanketed stellar flux
resulting from this analysis is then used as input for the photo-ionization model of the
PN (Sec.~\ref{sec:nebular_modelling}).

\subsection{The Near-IR, Optical and UV Stellar Spectra}
\label{ssec:the_stellar_spectrum}

The spectrum of SwSt~1 contains a mixture of stellar and nebular emission lines.
The stellar lines are weaker and fewer than for the [WC10] central stars
CPD--56~80$^{\rm o}$8032 and He~2-113 (DBS97), while the nebular lines
are stronger. The nebular Balmer lines,
the lines of He~{\sc i}, and, to a lesser extent, those of C~{\sc ii} 
and O~{\sc ii}, could be blends of stellar and nebular emission. 
In particular in the UV C~{\sc iii}] $\lambda$1909 is a blend of stellar and nebular 
emission, with the nebular emission contributing about 40\% of the flux to this 
feature (Fig.~\ref{fig:iue}).
The C~{\sc ii}] $\lambda$2326 feature is thought to be mostly of nebular origin, although this cannot
be checked since only low resolution IUE data exists for this wavelength range.
The UV C~{\sc iii}-{\sc iv} and Si~{\sc iv} emission lines are entirely of stellar origin.

In Sect.~\ref{sssec:the_balmer_lines}, we will show
that it is unlikely that hydrogen is present in the star, so that
the main contributors to the Balmer lines are likely to be nebular hydrogen and,
to a much lesser extent, stellar He~{\sc ii}. Additionally,
there is no evidence of a significant stellar contribution to the strong
nebular He~{\sc i} lines. Except for
$\lambda$5876, which exhibits a broad pedestal at the base of the nebular component,
other strong He~{\sc i} lines at 3889, 4471, 4713 and 6678~\AA , show no evidence of having 
a stellar component. The broad pedestal at the base of He~{\sc i}
$\lambda$5876 is likely to belong in part to
C~{\sc iii} M~20, with components at (in order of intensity)
5894.07, 5871.69, 5858.35, 5880.54, 5863.24 and 5872.10~\AA , since 
the four strongest components are 
observed at approximately the predicted relative strengths. 
In the near-IR we observe He~{\sc i} at 1.083~$\mu$m, although the lower resolution
of this spectrum does not allow us to distinguish between nebular and stellar 
components to this line. Stellar
He~{\sc ii} $\lambda$4686 is clearly visible in 
emission, He~{\sc ii} $\lambda$5411 has a weak P-Cygni absorption
but its emission, if present, is blended with the nebular [Fe~{\sc iii}] line
at 5412.0~\AA ;
He~{\sc ii} $\lambda$4541 has a weak emission to a blue-shifted absorption.

C~{\sc iii} lines dominate over C~{\sc ii} lines in SwSt~1's spectrum, indicating a higher degree of 
ionization than for the two [WC10] stars analyzed by DBS97. In the
optical we observe C~{\sc iii} $\lambda$4650,
$\lambda$5696, $\lambda$8500.
$\lambda$8500 is a blend of three lines: the stellar C~{\sc iii} $\lambda$8500,
a strong nebular line with radial velocity-corrected wavelength at 8502.1~\AA , 
likely to be itself a blend of H~{\sc i} Paschen~16 and [Cl~{\sc iii}] $\lambda$8500
([Cl~{\sc iii}] $\lambda$8481 is also present), 
and a weaker line at 8499.7~\AA . An unidentified nebular line at this
spectral position was also observed by Keyes et al. (1990) in the spectrum 
of NGC~7027. In the near-IR SOFI spectrum we observe
C\,{\sc iii} at 9710~\AA, and at 1.198~$\mu$m. 

\begin{figure}
\vspace{7cm}
\includegraphics{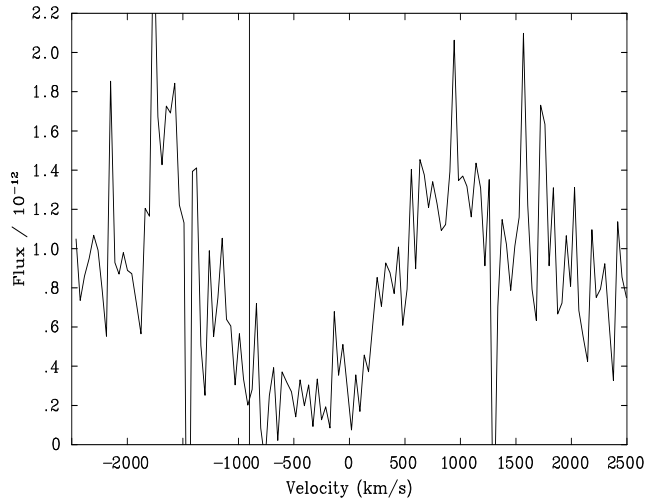}
\caption{The C~{\sc iv} resonance doublet profile in the IUE high resolution spectrum (SWP17068HL). 
The x-axis has been converted to velocity
space; v=0~km/s corresponds to 1548.2~\AA , the bluer of the
two components of the doublet (at 1548.2 and 1550.8~\AA ). 
The vertical line marks the chosen v$_{\rm black}$=--900~km~s$^{-1}$.}
\label{fig:1550}
\end{figure}

C~{\sc iv} $\lambda$$\lambda$5801,12, $\lambda$5470 and $\lambda$4441 are observed, albeit weakly,
while in the UV we see C~{\sc iv} $\lambda$$\lambda$1548,1551.
C~{\sc ii} lines are observed at 4267~\AA , 6578~\AA\ and 7231.3~\AA\ (blended with telluric 
absorptions). All C~{\sc ii} features exhibit stellar as well
as nebular components.
Semi-forbidden C~{\sc ii}] $\lambda$$\lambda$2324.2,2325.4 (likely to be entirely of nebular origin)
and C~{\sc iii}]
$\lambda$$\lambda$1906.7,1908.7 (having a nebular as well as a stellar component - Fig.~\ref{fig:iue}), 
are also observed. 

O~{\sc iii} $\lambda$5592 is observed in emission with a weak P-Cygni absorption, while
the O~{\sc ii} stellar spectrum is very weak (e.g. $\lambda$$\lambda$4591.0,4596.0). 
No nitrogen lines are observed, although there is a possibility
that a feature observed in the IUE high resolution spectrum at 1718.92~\AA\
is indeed N~{\sc iv} $\lambda$1718.5 and not Si~{\sc iv} $\lambda$1722.5. 
The optical lines of nitrogen used by Leuenhagen \& Hamann (1998) 
(i.e. N~{\sc iii} $\lambda$$\lambda$4097,4104) to determine an upper limit 
for the
abundance of nitrogen are not observed in our spectrum. 
N~{\sc ii} $\lambda$3995 and $\lambda$4682 are also unobserved.
The N{~\sc v} line marked by Flower et al. (1984) in the 
UV spectrum at 1438.8 and 1442.8~\AA\ 
is almost certainly C~{\sc iii} $\lambda$1347.4.            
Si~{\sc iv} $\lambda$$\lambda$1393.7,1402.8 are observed in the UV spectrum, 
as are weak optical lines at
4089 and 4116~\AA\ (M1).

From the high resolution
IUE spectrum of the C~{\sc iv} $\lambda$1550 line we measure a $v_{\rm black}$
(which corresponds to the wind terminal velocity $v_\infty$; Prinja, Barlow \& Howarth 1990) 
of (900 $\pm$ 150)~km~s$^{-1}$ (Fig.~\ref{fig:1550}). 
We notice how this value is much larger than anything implied by
the optical line widths and P-Cygni features (which is why Leuenhagen \& Hamann (1998) adopted
a much lower value for $v_\infty$ of 400~km~s$^{-1}$). This is likely to be
a consequence of a particularly low wind density.

\subsubsection{The Balmer Lines}
\label{sssec:the_balmer_lines}

\begin{figure*}
\vspace{10cm}
\includegraphics{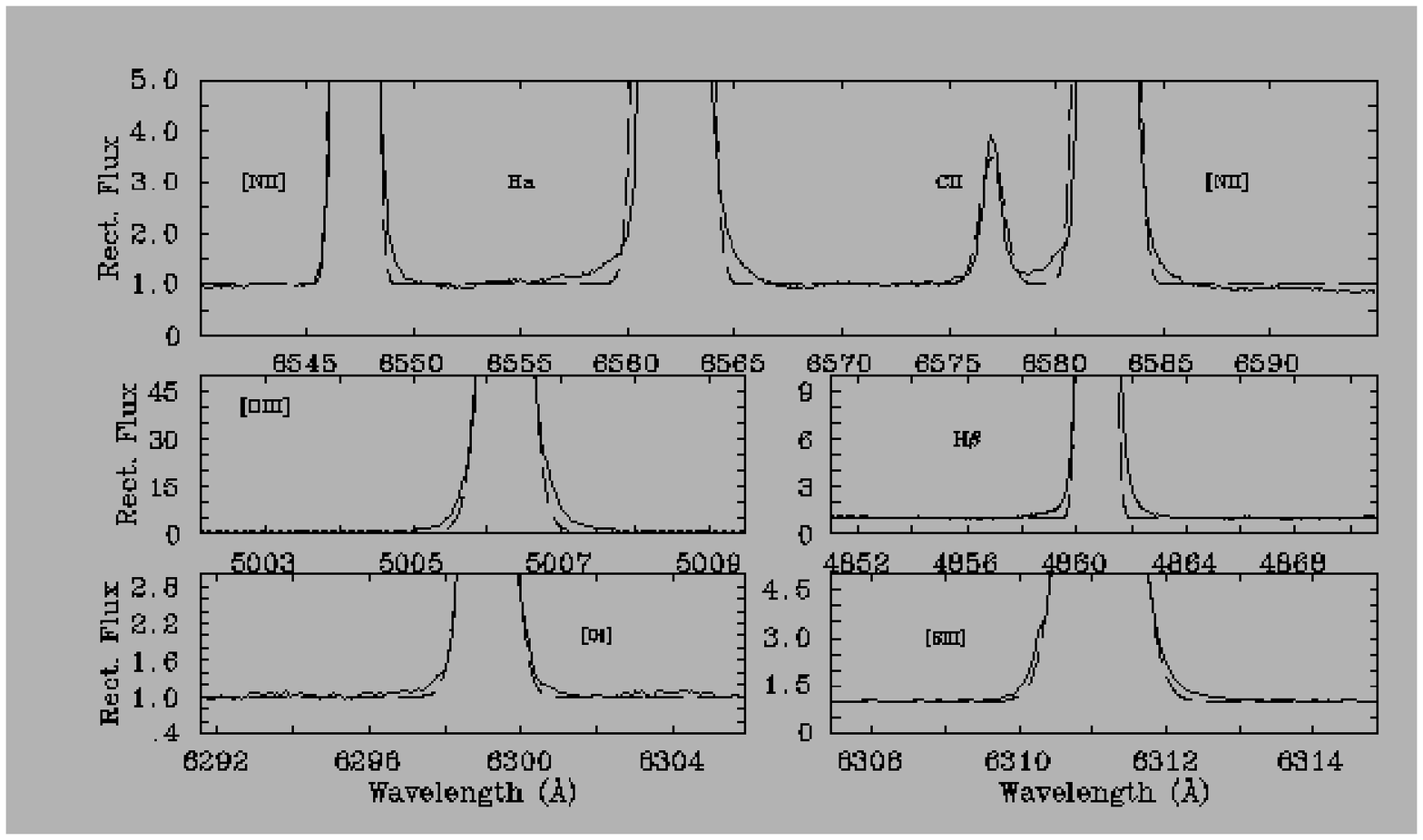}
\caption{Gaussian fits to nebular lines, showing the nature of the
asymmetry at their bases to be of nebular or instrumental origin.}
\label{fig:balmer_shapes}
\end{figure*}

DBS97 noted that the Balmer line shapes of the PN around the two [WC] central
stars CPD--56$^{\rm o}$8032 and He~2-113 have a broad pedestal. Leuenhagen et al. (1996)
determined an upper limit for the {\it stellar}
hydrogen abundance by fitting this
pedestal with synthetic profiles from their non-Local Thermodynamic
Equilibrium (non-LTE) code. 
However DBS97 pointed out
that the same broad pedestal was present at the base of
{\it forbidden} lines which are entirely of nebular origin. This
indicates that either a complex nebular velocity field is the cause
of the
irregular nebular line shapes and of the pedestals at the base of the
Balmer
lines, or perhaps more likely, that the pedestals observed at the
base of these strong lines is due to the
grating instrumental profile. 

Since the presence of hydrogen in cool [WC] central stars would give a clue about 
their origin, we have investigated the nebular lines of SwSt~1. In
Fig.~\ref{fig:balmer_shapes} we present Gaussian fits to H$\alpha$ and
H$\beta$, along with other forbidden lines.
Although the Balmer lines do present a base which is broader than the
Gaussian fit, many other
nebular lines, i.e. [N~{\sc ii}] $\lambda$$\lambda$6548,6583 and
[O~{\sc iii}] $\lambda$5007, have a similar shape. 
There is therefore no conclusive evidence for the presence of
stellar hydrogen.
In Sect.~\ref{ssec:stellar_model_results} we demonstrate how a hydrogen abundance
of 10\% (by mass) as determined by Leuenhagen and Hamann (1998) is not inconsistent with 
our own fits.

\subsection{Description of the Model Codes}
\label{ssec:description_of_the_model_code}

The code adopted to carry out the stellar analysis
is the improved Sobolev approximation code {\sc isa}-Wind (de~Koter et al. 
1993, 1997). This modelling program solves the radiation transfer equation under the
condition of statistical equilibrium in a spherical, extended atmosphere. The line
radiation transfer is implemented using an improved version of the Sobolev approximation,
which has been shown to be appropriate for the thick atmospheres of 
Wolf-Rayet stars (de~Koter et al. 1997, Crowther et al. 1999 and De Marco et al. 2000).

The wind of SwSt~1, however, is less dense than the winds of most Wolf-Rayet stars, 
as can be inferred from the overall weakness of its spectral lines.
This indicates that the inner, slow moving parts of the stellar atmosphere are exposed
in this region, and the Sobolev approximation might break down. 
To check on the results obtained with {\sc isa}-Wind,
we used the co-moving frame code {\sc cmfgen} (Hillier \& Miller 1998). 
{\sc cmfgen} confirmed the results obtained with
{\sc isa}-Wind. 
However, some of the differences between the two codes are instructive, and although
in what follows we refer mostly to {\sc isa}-Wind, results
obtained using {\sc cmfgen} are presented alongside.

Line blanketing is included in the {\sc isa}-Wind model by a Monte-Carlo 
approximation, which includes millions of spectral lines.
This is described in detail by Schmutz et al. (1991). Comparison tests have been
carried out (Crowther et al. 1999, De Marco et al. 2000)
and show the technique to compare reasonably well to a detailed
line-blanketing calculation carried out with {\sc cmfgen} (Hillier \& Miller 1998), although the flux
shortwards of 400~\AA\ was too hard in the Monte-Carlo technique.

\subsubsection{The Model Atom}
\label{sssec:the_mdel_atom}

The number of atomic levels included in the calculation influences the
line strengths. Ideally one would want to employ as many lines as possible
of as many elements as possible. Unfortunately the larger the atomic model
the slower the calculation. As a result a compromise has to be adopted.
The atomic model used by our {\sc isa}-Wind calculation uses Opacity Project
data (Seaton 1995) and is
summarised in Table~\ref{tab:atomic_model}. The use of super-levels (SL)
reduces computational times while at the same time including high-lying
levels. Super-levels are groups of levels within which populations are
calculated in LTE. The grouping scheme changes for different ions. In
Table~\ref{tab:atomic_model} we indicate the scheme used: ``n=4'' means
that all levels up to and including principal quantum number 4 are
included. ``sing/trip'' means that singlets and triplets are grouped
separately; ``thr.'' indicates that enough super-levels were constructed
to include all levels listed in the Opacity Project database up to the
ionization threshold.  The effect of slight variations in our grouping
scheme (as well as the total number of SL) on the synthetic lines was
studied for the cases of C~{\sc ii} and C~{\sc iii} and found not to alter
the synthetic lines. Only
helium, carbon, oxygen, silicon and iron are all included explicitly in
the {\sc cmfgen} calculation, but in somewhat greater detail than in the 
case of {\sc isa}-Wind.

\begin{table}
\caption{Summary of the atomic model adopted in our {\sc isa}-Wind non-LTE simulations
of SwSt~1's stellar wind.}
\begin{tabular}{lrlrlr}
Ion           & \#   & Upper   & \# & SL Grouping & Dielec-\\
              &Levels& Level   & SL      &             & tronic\\
H~{\sc i}     &  15  & --         & -- & -- &--\\
He~{\sc i}    &  19  & $^1$P n=4  & 32 & n=20 sing/trip &--\\
He~{\sc ii}   &  20  & --         & -- & -- &--\\
C~{\sc ii}    &  36  & $^2$G     n=3 & 2  & thr. doub/quart &26\\
C~{\sc iii}   &  31  & $^1$S     n=5 & 8  & thr. sing/trip &31\\
C~{\sc iv}    &   9  & $^2$F     n=4 & 3  & n=7 &--\\
N~{\sc i}     &   3  & $^2$P$^o$ n=2 & -- & -- &--\\
N~{\sc ii}    &  29  & $^3$S     n=2 & -- & --& --\\
N~{\sc iii}   &  11  & $^2$D     n=3 & 3  & n=4 doub/quart&--\\
N~{\sc iv}    &  12  & $^1$D     n=3 & -- & -- &40\\
N~{\sc v}     &  14  & $^2$D     n=6 & -- & -- &--\\
O~{\sc i}     &   2  & $^3$S$^o$ n=3 & -- & -- &--\\
O~{\sc ii}    &   3  & $^2$P$^o$ n=2 & -- & -- &--\\
O~{\sc iii}   &  41  & $^5$S$^o$ n=3 & -- & -- &--\\
O~{\sc iv}    &  11  & $^2$D     n=3 & 3  & n=4 doub/quart &--\\
O~{\sc v}     &  12  & $^1$D     n=3 & -- & -- &34\\
Si~{\sc ii}   &   7  & $^2$P     n=4 & -- & -- &--\\
Si~{\sc iii}  &  16  & $^3$F$^o$ n=4 & 2  & n=5 sing/trip &--\\
Si~{\sc iv}   &  12  & $^2$G     n=5 & 1  & n=6 &--\\
S~{\sc iii}   &  11  & $^3$P$^o$ n=3 & -- & -- &--\\
S~{\sc iv}    &   7  & $^2$S$^e$ n=4 & -- & -- &--\\
S~{\sc v}     &   7  & $^1$S     n=3 & -- & -- &--\\
S~{\sc vi}    &   7  & $^2$F$^o$ n=4 & -- & -- &--\\
\end{tabular}
\label{tab:atomic_model}
\end{table}

\subsection{Photon Loss}

``Photon loss" is the name given by Schmutz (1997) to the
interaction between the photon field of the He~{\sc ii}
Ly$\alpha$ line at 303.78~\AA\
and nearby metal lines (e.g. the O~{\sc iii} Bowen lines at
305.72 and 303.65~\AA ). This can be responsible for a decrease
in the ionization equilibrium of the wind with respect
to the equilibrium obtained when no such
line-line interaction is accounted for.
In order to fit the observed spectrum when photon loss is included in the
model, the model's stellar temperature
has to be higher, which results in a higher
spectroscopic luminosity. This contributes to
a lower wind performance number and allowed Schmutz (1997) to calculate a
velocity structure for the outer wind of WR6.

De Marco et al. (2000) implemented this approximation in the Sobolev code
{\sc isa}-Wind and showed that results were similar to those obtained using the
code {\sc cmfgen} (for the WC8 stellar component to the $\gamma$~Velorum binary), 
where line-line interaction is taken into consideration explicitly.
Here, we have used the ``photon-loss" option, which
resulted in only a slightly higher effective temperature
($\sim$2000~K). The effect has been
shown by Schmutz (1998) not to be important in the case of O stars, which
have
weaker winds. This also seems to be the case for the weaker wind of SwSt~1.
{\sc cmfgen} naturally includes line-line interaction and therefore ``photon-loss"
is automatically taken into account (De Marco et al. 2000).

\subsection{Stellar Modelling Technique}
\label{ssec:stellar_modelling_technique}

Traditionally, modelling of WR stars starts with matching a subset of the observed
He~{\sc i} and He~{\sc ii} lines with model line profiles, from which
approximate values for the effective
temperature and mass-loss rate of the star are determined.  
The metal abundances are then varied, to afford fits to a representative sample of
their lines.
Even in modern calculations the velocity law is seldom varied from either a simple or a composite
beta-law (e.g. Hamann \& Koesterke 2000).

A weakness of this technique is that in a significant number of cases,
the choice of diagnostic lines
determines what parameters are derived. If excellent fit
quality can be obtained for a large number of lines, this gives
confidence in the derived parameters, but if
only a small subset of the observed lines can be fit by any given model,
the parameters are only as good as the choice of diagnostics.
Additionally, running non-LTE line-blanketed codes is very time consuming, so that
once a reasonable match to a set of lines is achieved, different parameter spaces are 
seldom explored to determine whether other parameter sets fit the emission lines
to a similar degree of accuracy.

The fast calculation speed of the {\sc isa}-Wind code allowed De Marco et al. (2000)
to show that similar synthetic spectra can sometimes 
be obtained with different combinations of temperature and mass-loss rate.
When only a poor fit quality is achieved, or when only a subset of the 
diagnostic lines is fit simultaneously, it might be hard to choose which
model best represents the star, leading to a degeneracy.

The spectrum of SwSt~1 presents a formidable challenge for spectral
synthesis, in that it is impossible to
fit all the diagnostic lines simultaneously. As a result, it is possible to 
derive different model parameters depending on which diagnostic lines we
decide to include. The exploration of a vast parameter space (stellar temperature,
mass-loss rate and metal abundance), made possible by our fast code, together with the use of
observational constraints not usually coupled with non-LTE analyses, made
it possible to
gain a reasonable confidence in our final choice of stellar parameters.
Our choice of diagnostic lines comprised all lines included within the levels of our
model atom, leaving out those between high-lying levels which might be under-populated.
For helium we used
He~{\sc ii} $\lambda$4686 ($\lambda$5411 
is blended with nebular a [Fe~{\sc ii}] line, and can only be used as an upper limit)
and He~{\sc i} $\lambda$5876,4471,6678. As pointed out above, there is extremely little
He~{\sc i} in this star, if any, so that the lines (severely blended with
nebular components) could only be used to impose an upper limit on
the strengths. To constrain the ionization equilibrium one has to appeal
to carbon:
C~{\sc ii} $\lambda$4267, C~{\sc iii} $\lambda$5696
and $\lambda$8500, and 
C~{\sc iv} $\lambda$$\lambda$5801,12 and $\lambda$5470 were employed. These lines are also used
to determine the carbon abundance.
To constrain the oxygen abundance
we used O~{\sc iii} $\lambda$5592 (all other lines are too weak or dominated by the nebula).

Once the diagnostic lines were chosen, an approximate value for the
temperature and mass-loss rate was obtained, for which reasonable fits
were achieved.  Since no combination of $T_*$, $\dot{M}$ and carbon
abundance fitted the spectrum to a satisfactory level, in that the
synthetic lines were always too broad in the core and too narrow at the
base, we experimented with different velocity laws. In particular we
tested velocity laws with progressively slower flows in the inner regions,
so as to reduce the line widths, maintaining the observationally-derived
value of $v_\infty$. The electron density change induced by the different
laws changes the wind opacity, with a repercussion on the line ionization
equilibrium of the star, and therefore on the line intensities.
Consequently different basic parameters are implied by different velocity
laws.

\subsection{Stellar Model Results}
\label{ssec:stellar_model_results}

\begin{figure*}
\vspace{22cm}
\includegraphics{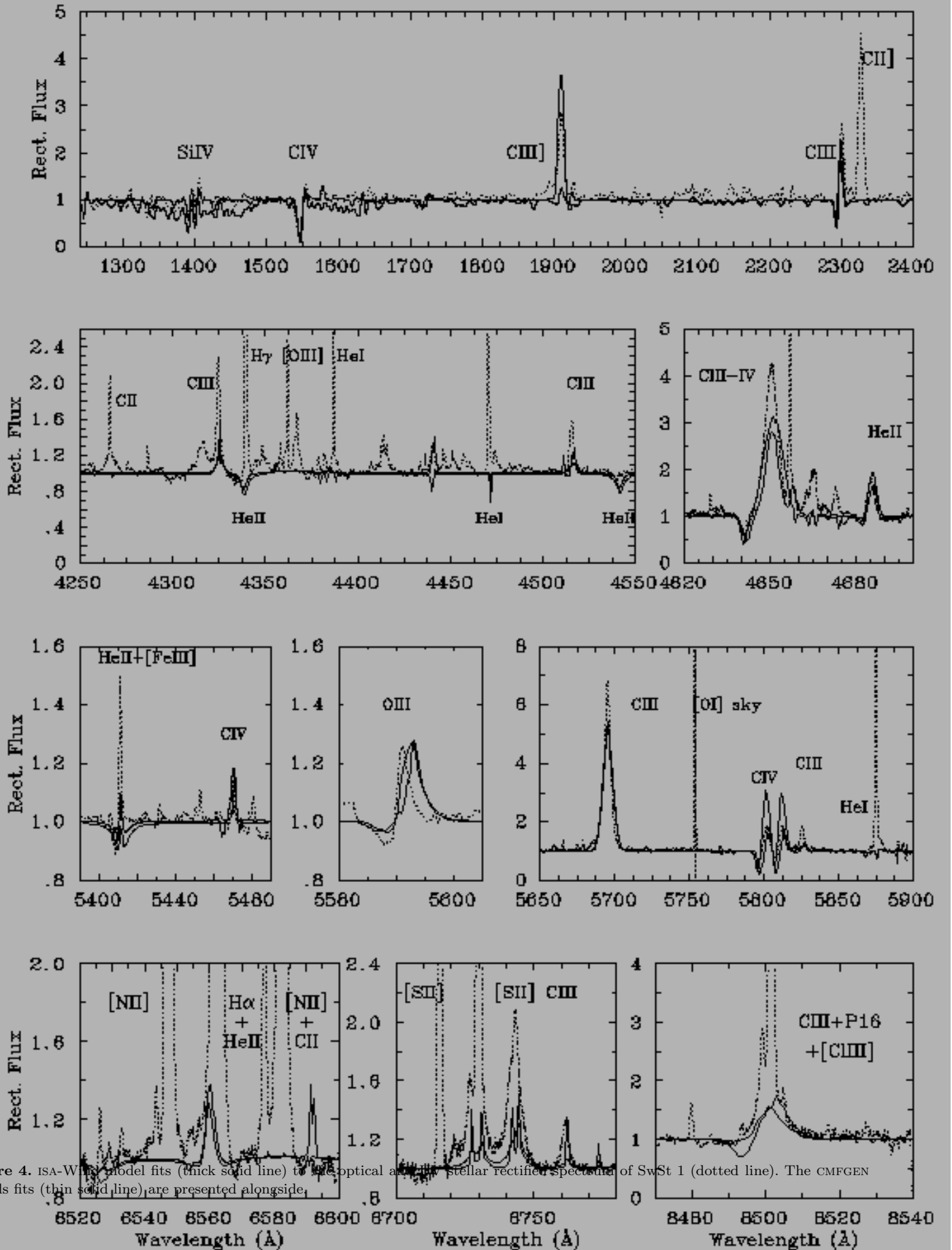}
\caption{{\sc isa}-Wind model fits (thick solid line) 
to the optical and UV stellar rectified spectrum of SwSt~1 (dotted line). 
The {\sc cmfgen} models fits (thin solid line) are presented alongside.
}
\label{fig:model_comp_final}
\end{figure*}

\begin{table*}
\begin{minipage}{180mm}
\caption{Stellar parameters from our non-LTE models ({\sc isa}-Wind and {\sc cmfgen}),
compared with those of Leuenhagen \& Hamann (1998; LH98), before
and after scaling them to our own adopted distance.}
\begin{tabular}{lllcrcccccccclcl}

V & E(B--V) & D & M$_V$ & T$_\ast$ & R$_\ast$ & log(L/& log($\dot{M}$/ & v$_\infty$ & $\eta$ & $\beta_H$&$\beta_{He}$&$\beta_C$&$\beta_O$ & $\beta_N$  & Ref.\\
(mag)& (mag)& (kpc)& (mag)& (kK)& (R$_\odot$)&  L$_\odot$) & M$_\odot$~yr$^{-1}$)& (km/s) &  & \% &\% & \% & \% & \% & \\

11.48& 0.46 & 2.0 & --1.45 & 40 &  2.0  & 3.95 & --6.72& 900 & 0.9& $<$10$^a$&37&51 & 12& $\sim$0 & {\sc isa}-Wind\\  
11.48& 0.48 & 2.0 & --1.51 & 46 &  1.5  & 3.95 & --6.84& 800 & 0.6&     0    &53&32 & 15&       0 & {\sc cmfgen}\\  
11.9 & 0.41 & 1.4 & --1.10 & 35 &  1.0  & 3.27 & --6.90& 400 & 1.3& $<$10 & 44 & 53 & 3 & $<$0.5 & LH98 \\
--   & --   & 2.0 & --1.45 & 35 &  1.4  & 3.49 & --6.79& 400 & 1.3& $<$10 & 44 & 53 & 3 & $<$0.5 & LH98 scaled\\
\noalign{\smallskip}
\multicolumn{14}{l}{$^a$ Limit from model profile fitting does not
include possibility of blend with nebular emission (see text).}\\
\end{tabular}
\label{tab:stellar_analysis}
\end{minipage}
\end{table*}

Model fits to the stellar spectrum of SwSt~1 are presented in 
Fig.~\ref{fig:model_comp_final} (thick and thin solid lines for {\sc isa}-Wind and {\sc cmfgen} models,
respectively, dotted line for the observations), while the model parameters are presented in
Table ~\ref{tab:stellar_analysis}, together with the results of Leuenhagen \&
Hamann (1998), scaled to our adopted distance. In what follows we discuss
the results obtained with the modelling code {\sc isa}-Wind, followed by a discussion of additional
results obtained with the code {\sc cmfgen} to later.

\subsubsection{{\sc isa}-Wind}

He~{\sc ii} $\lambda$4686
is well reproduced, while He~{\sc ii} $\lambda$5411 and He~{\sc i} 
$\lambda$$\lambda$5876,6678,
and other lines which suffer blending with nebular emission,
are within reasonable limits.
Other synthetic He~{\sc ii} lines (e.g. $\lambda$4542 and $\lambda$4200)
are in absorption with wide wings. Since these  wings are not observed,
and He~{\sc i} lines are mostly of nebular origin, we have have not
used these lines as diagnostics.

C~{\sc iv} $\lambda$5801,12 is over-predicted, C~{\sc iv} $\lambda$5470 is
well matched, while C~{\sc iv} $\lambda$4441 is somewhat weak. In order to
fit C~{\sc iv} $\lambda$$\lambda$5801,12 the effective temperature
could not
be higher than 35~kK, no matter what other parameters are adopted. However
at this temperature the $\lambda$5470 line is in weak absorption. C~{\sc
iv} $\lambda$5470 has a better-understood formation mechanism compared to
$\lambda$$\lambda$5801,12 (Hillier \& Miller 1999, Dessart et al. 2000)
and was therefore given priority in the fitting procedure. Additionally, 
at
35~kK the fit to the C~{\sc iii} $\lambda$4650 line is much worse, with
the synthetic line too weak. C~{\sc iii} $\lambda$5696,8500 are also well
matched by the model but C~{\sc ii} $\lambda$4267 is poorly matched, with
the
model line being much too weak. However the C~{\sc ii}
$\lambda$$\lambda$6578,6583 blend has a very reasonable fit 
(C~{\sc ii} $\lambda$$\lambda$7231,7236 is severely affected by telluric absorption
and blending with nebular emission line components and was therefore not fitted).  The problem
with C~{\sc ii} $\lambda$4267 might be linked to insufficient levels in
the atomic models. However we have experimented with more levels and
included a considerable number of super-levels and dielectronic
recombination lines.  Since a much stronger synthetic $\lambda$4267
(obtained by lowering the effective temperature) would lead to
over-fitting $\lambda$$\lambda$6578,83, we still suspect the model atom is
the reason for the problem. If a cooler stellar temperature is used to
try and enhance C~{\sc ii} $\lambda$4267, all fits deteriorate, even if
the mass-loss rate is reduced to restore the He~{\sc i}/He~{\sc ii}
equilibrium. Increasing the mass-loss rate so as to enhance recombination
of C~{\sc iii} to C~{\sc ii}, and so increase the strength of the C~{\sc
ii} $\lambda$4267 line, also fails. A higher mass-loss rate is also
excluded by the fact that by increasing its value by 0.1~dex from the
value reported in Table~\ref{tab:stellar_analysis}, all model lines of
C~{\sc iii} acquire deep P-Cygni absorptions, which are not observed.
Additionally the He~{\sc i} lines increase in strength, in a way that
cannot be compensated for by a larger temperature without losing the
C~{\sc iii}/C~{\sc iv} balance.

Increasing the carbon abundance, in steps, 
from C/He=0.2 to 3 (by number), while adjusting other parameters, does not 
achieve an improvement in the fits. On the other hand, once the effective 
temperature and mass-loss rate have been determined, the C/He ratio can be
determined with a 50\% accuracy. 
Even with the highest carbon abundance, the
C~{\sc ii} line did not increase for the temperature regime
which satisfied the other lines. Additionally, when enhancing the
C/He ratio the ionization equilibrium of the star is upset in
such a way that He~{\sc ii} lines are drastically decreased, which
results in having to increase the temperature, at which point other
carbon lines do not fit any more, most noticeably C~{\sc iv} $\lambda$$\lambda$5801,12,
whose excitation is almost entirely dependent on temperature. 

\begin{figure*}
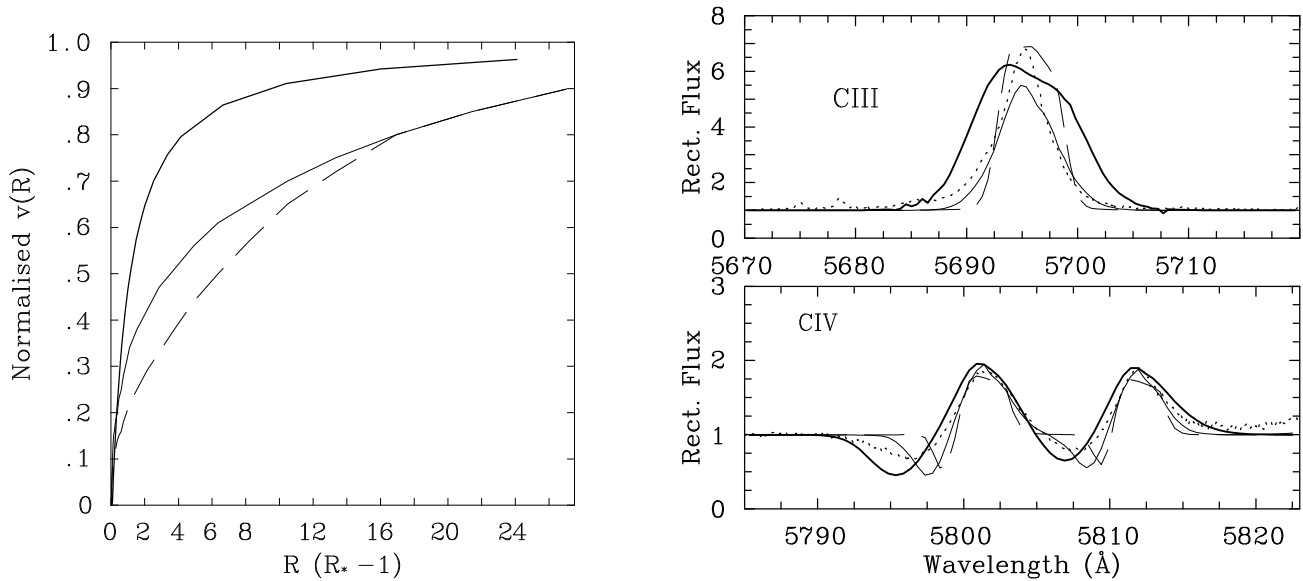

\vspace{8cm}
\includegraphics{velocity_laws.eps}
\includegraphics{velocity_law_profiles.eps}
\caption{{\bf Left panel}: alternative velocity laws compared to a $\beta$=1 law 
(thick solid line). The adopted law (thin solid line) significantly improved
the line fits, making lines broad at the base and narrow in the core, and 
reproduced the position of their P-Cygni absorptions, 
while the other laws produced lines which were too broad or too narrow.
{\bf Right panel}: {\sc isa}-Wind profile fits to C~{\sc iii} $\lambda$5696 (top) and
C~{\sc iv} $\lambda$$\lambda$5801,12 (bottom).
The dotted line represents the data, the thick solid line represents the model calculated
using a $\beta$=1 velocity law, the thin solid line represents the model with the velocity law
used for the final model of this paper, and the dashed line shows the 
slowest velocity law attempted. Model parameters are adjusted to fit the 
peak intensity of C~{\sc iv} $\lambda$$\lambda$5801,12.}
\label{fig:velocity_laws}
\end{figure*}

Returning to the C~{\sc iii} line blend at 4650~\AA , and to line shapes
in general, we have used a velocity law which is slower than $\beta$=1 in
the inner parts of the wind (in accordance with that found by Hillier \&
Miller (1998), Crowther et al. (1999) and De Marco et al. (2000)). For
every parameter set, we tested a number of velocity laws with
progressively slower winds in the inner parts of the flow. These are shown
in Fig.~\ref{fig:velocity_laws}. The slower the velocity law, the stronger
and narrower the lines become (for identical stellar parameters). As a
result each adopted velocity law implied a new set of parameters,
resulting in a trend of a somewhat cooler stellar temperature and lower
mass-loss rate for slower velocities. For the slower velocity laws,
narrower line profiles were accompanied by more redward-displaced P-Cygni
absorptions. As a result it was relatively easy to choose the best
velocity law from the position of the P-Cygni absorption (as well as the
overall line width
and shape) of lines such as C~{\sc iii} $\lambda$4650 and C~{\sc iv}
$\lambda$$\lambda$5801,12. The velocity law which matched the position of
the P-Cygni profiles is marked as a thin solid line in
Fig.~\ref{fig:velocity_laws} (left panel) and can be parametrised
following the formula of Hillier and Miller (1998) with $v_{\rm ext}$ =
500~km~s$^{-1}$, $v_\infty$ = 900~km~s$^{-1}$, $\beta_1$=1 and
$\beta_2$=8. This resulted in the best match to the line shapes. In
Fig.~\ref{fig:velocity_laws} (right panel) we present profile fits
obtained using a $\beta$=1 law, and two slower laws. The model parameters
were adjusted in each case to reproduce the line's peak intensity. As can
be seen, the $\beta$=1 law generates profiles which are too broad, while
the dot-dash line produces profiles too narrow at the base but too broad
in the core.

In summary, although the fits presented in Fig.~\ref{fig:model_comp_final} (thick solid line)
are not excellent, there is no other parameter combination of mass-loss
rate, temperature and C/He number ratio\footnote{The parameter space
explored spanned from 30kK to 45kK in stellar temperature,
$\log({\dot{M}}/$M$_\odot$)=--6.0 to --6.7 and C/He=0.2 to 3.0.} which
would satisfy the observed spectrum to a higher degree of accuracy. The
derived quantities and their errors are therefore the following:
$T_*$=(40\,000 $\pm$ 2000)~K, $\log({\dot{M}}/$M$_\odot$) = (--6.72 $\pm$ 0.1),
C/He = (0.46 $\pm$ 0.20) and O/He = (0.08 $\pm$ 0.02) (by number). The wind
performance number $\eta$ (=$L/({\dot{M}}{v_\infty}$c) is 0.9, indicating
that the wind
of SwSt~1 can
be driven radiatively in the single scattering limit. The derived stellar
luminosity of 8900~L$_\odot$ is towards the high end for a PN central star
(implying a relatively large stellar mass).  It is difficult to determine
the mass which corresponds to this luminosity since the highest
core mass for which Vassiliadis and Wood (1994) calculated Milky Way 
initial abundance helium-burning
tracks was 0.63~M$_\odot$, but inspection of their Fig.~8 indicates
that our derived luminosity for SwSt~1 at 40\,000~K is only slightly above
the LMC-abundance evolutionary track for a 0.679~M$_\odot$ core mass.

The helium, carbon and oxygen abundances are 37\%, 51\% and 12\% by mass, respectively,
implying a C/O mass ratio of 4.3. The oxygen abundance in particular is obtained
from fitting the O~{\sc iii} line at 5592~\AA .
The lower and upper limits for the O/He
ratio allowed by the fit of the $\lambda$5592 line are 9\% and 15\%.
Additional uncertainty derives from fitting a single
line. We note that if a lower effective temperature were to be adopted,
the oxygen abundance determined by a fit to this O~{\sc iii} line would be
higher still.

In Fig.~\ref{fig:model_flux} we show the {\sc isa}-Wind synthetic spectrum, compared with the
IR, optical and IUE de-reddened data. A slight over-compensation of the 
2200~\AA\ feature by the adopted E(B--V)=0.46 is not considered worrying
in view of the known variations in the relative strength of this
feature (Mathis 1994),
since the overall continuum slope is well matched by our V-scaled
synthetic spectrum 
(if we make an exception for the far blue and red edges
of the optical spectrum). By adopting reddening values as low as 0.30 (which
nulls the 2200~\AA\ feature) the continuum slope is clearly too gentle
to match the model, while for a reddening value of 0.50 the slope will be too
steep. In Fig.~\ref{fig:model_flux} we also plot the near-IR SOFI spectrum (scaled
to the J magnitude of Allen \& Glass (1974)) and colour-corrected
IRAS photometry
values, together with blackbody curves with temperatures
1200~K and 230~K. These curves indicate 
that some of the dust around this PN is very hot. This is in
agreement with the study of Zhang \& Kwok (1991).

\subsubsection{{\sc cmfgen}}

The fit quality obtained with the code {\sc cmfgen} is very similar
to that obtained with {\sc isa}-Wind. Although some lines are better 
reproduced (e.g. C~{\sc iv} $\lambda$$\lambda$5801,12) others are
somewhat worse (e.g. C~{\sc iv} $\lambda$4441). The broad He~{\sc ii}  
absorption lines wings
obtained with {\sc isa}-Wind are somewhat narrower
in the models fits of {\sc cmfgen}, in better agreement with the data.
C~{\sc iii} $\lambda$1909 is weaker and more in agreement with the observations
(note that in Fig.~\ref{fig:model_comp_final} this line is a blend of
stellar and nebular components, as can be appreciated by inspecting Fig.~\ref{fig:iue}).
On the other hand the shape of the C~{\sc iii}-{\sc iv} blend at $\sim$4650~\AA\
is somewhat worse in the {\sc cmfgen} fits, and the C~{\sc iii} line at 8500~\AA\ has
an unobserved P-Cygni absorption. The fit to the only oxygen diagnostic line,
i.e. O~{\sc iii} $\lambda$5592, is similar to the one obtained with {\sc isa}-Wind.
The stellar parameters obtained with {\sc cmfgen} are presented alongside those
obtained with {\sc isa}-Wind in Table~\ref{tab:stellar_analysis}. 
Overall, the derived stellar parameters for SwSt1 using CMFGEN match
closely with those from ISA-wind, except that the stellar radius is
smaller, implying a higher stellar temperature of 46kK.       
The mass-loss listed in 
Table~\ref{tab:stellar_analysis} does not consider clumping. Almost identical
fits can be obtained with a filling factor of 10\% with clumping starting at 
$v=100$~km~s$^{-1}$. In this case the mass-loss is 7$\times$10$^{-8}$M$_\odot$~yr$^{-1}$
or a factor of two lower than obtained without clumping. Due to the relatively
low wind density, the electron scattering wings
cannot be used to constrain the clumping. The only noticeable changes are in 
C~{\sc iii} $\lambda$9710 and $\lambda$1909 where clumped models give
20\% weaker emission. However, since the latter is dominated  by the nebula and
the former is only included in our low resolution SOFI data, these cannot provide any useful
constraint.

The only considerable difference between the modelling efforts with {\sc isa}-Wind and
{\sc cmfgen} is that the stellar abundances are discrepant (Table~\ref{tab:stellar_analysis}). This is 
possibly due to the different cocktail of atomic data used by the
model codes, although, as it is frequent with complex non-LTE codes, it might also be 
due to a mix of factors hard to trace. In particular the carbon and oxygen mass fractions are
32\% and 15\%, respectively, compared to {\sc isa}-Wind's 51\% and 12\%. This
translates into a C/O mass ratio of 2.1  for {\sc cmfgen} versus 4.3  for {\sc isa}-Wind.
Both {\sc cmfgen} and {\sc isa}-Wind values are higher than solar ($\sim$0.4). The maximum value 
that the C/O ratio can attain according to the models (Herwig 1999), is the intershell value, $\sim$4.
So, although both values are acceptable, the {\sc isa}-Wind value would imply an extremely
effective third dredgeup. 

We obtain a good fit of the {\sc cmfgen} model flux distribution to the dereddened data, 
although a slightly higher
value of the reddening (E(B--V)=0.48~mag) had to be adopted.

\subsubsection{Comparison with the Analysis of Leuenhagen \& Hamann 1998}

Leuenhagen \& Hamann (1998) present fits to their 1~\AA -resolution
optical spectrum that 
are not dissimilar in quality from our own, although they fit a smaller spectral
range. The effective temperature of 40~kK derived with the {\sc isa}-Wind modelling code, is higher than their
35~kK. Our mass-loss rate is 
similar to the one derived by Leuenhagen \& Hamann (once the different 
distances are taken into account),
while our final wind velocity is more than twice as large.
Leuenhagen \& Hamann (1998) have not determined $v_\infty$ from the C~{\sc iv} line,
limiting themselves to fitting the widths of the optical lines. 
We have already seen how
these lines are peculiarly narrow and in apparent disagreement with the value indicated
by the P-Cygni absorption of C~{\sc iv} $\lambda$1550. Inspecting their
Fig.~3 we notice that their model optical lines are too narrow and suffer from the same 
problem as our fits with a $\beta$=1 velocity law. 

The hydrogen abundance of 10\% (by mass) determined by Leuenhagen \&
Hamann (1998) as an upper limit, when used in our model, produces a
synthetic H$\alpha$-He~{\sc ii} line blend
which is not entirely in disagreement with our observations.
Therefore, based {\it solely} on this fit, we confirm their determination.
On the other hand, as already noted in Sec.~\ref{sssec:the_balmer_lines}
the blend's shape might be due to either nebular emission or the
instrumental profile, since nebular forbidden lines exhibit the 
same pedestal. Our {\sc isa}-Wind carbon abundance is
similar to that determined by Leuenhagen \& Hamann (1998), while the
carbon abundance obtained with {\sc cmfgen} is lower. Our
oxygen abundance, obtained with either of the two codes, is higher
than the abundance derived by Leuenhagen \& Hamann (1998). 

\begin{figure*}
\vspace{14cm}
\includegraphics{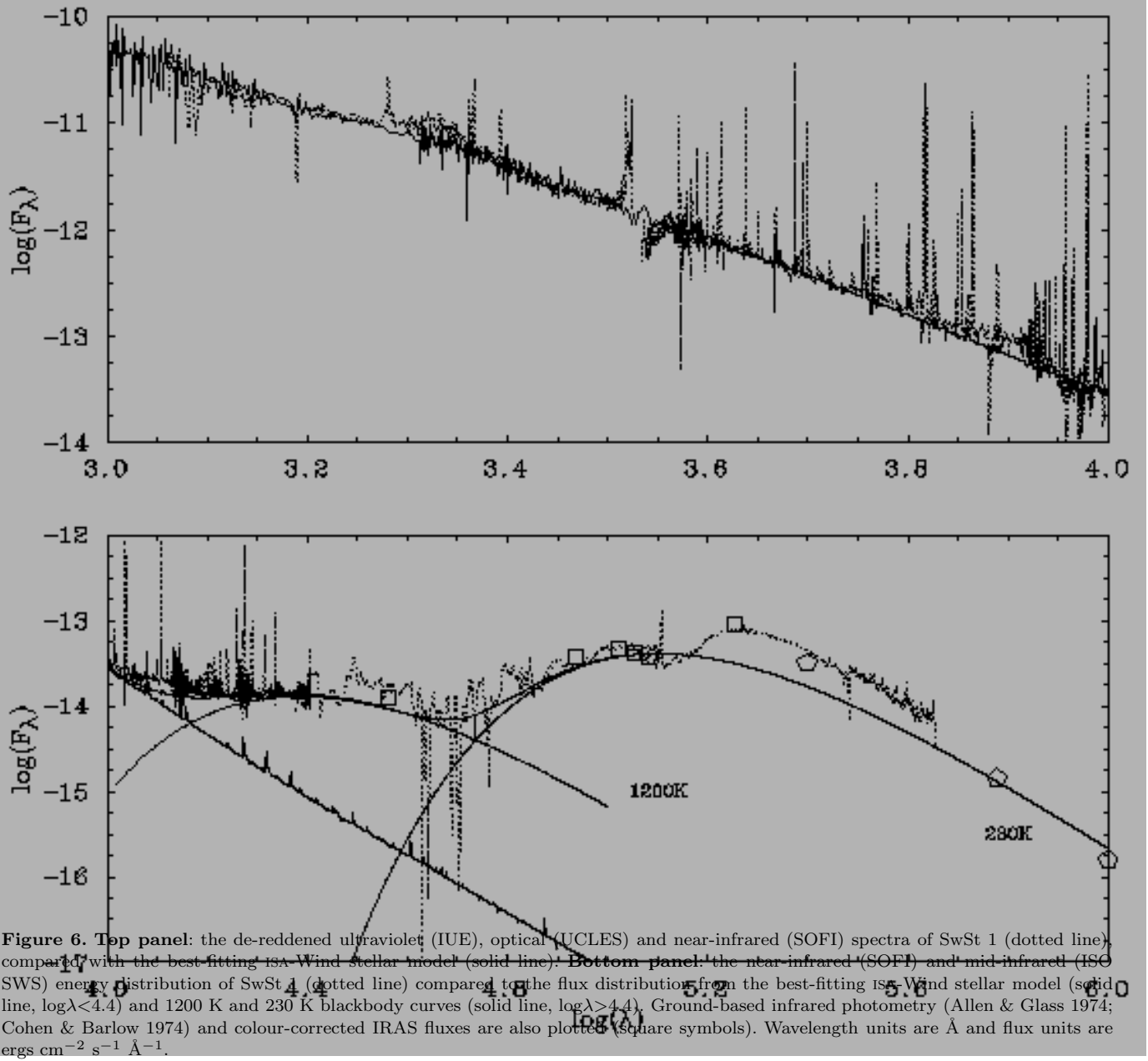}
\caption{{\bf Top panel}: the de-reddened ultraviolet (IUE), optical (UCLES) and
near-infrared (SOFI) spectra of SwSt~1 (dotted line), compared with the best-fitting {\sc isa}-Wind
stellar model (solid line). {\bf Bottom panel}: the near-infrared (SOFI) and mid-infrared (ISO SWS)
energy distribution of SwSt~1 (dotted line) compared to the flux
distribution from the best-fitting {\sc isa}-Wind
stellar model (solid line, $\log$$\lambda$$<$4.4) and 1200~K and 230~K 
blackbody curves (solid line, $\log$$\lambda$$>$4.4).
Ground-based infrared photometry (Allen \& Glass
1974; Cohen \& Barlow 1974) and colour-corrected IRAS fluxes are also
plotted (square symbols).
Wavelength units are \AA\ and flux units are
ergs~cm$^{-2}$~s$^{-1}$~\AA $^{-1}$.}
\label{fig:model_flux}
\end{figure*}

\section{Nebular Abundance Analysis}
\label{sec:nebular_abundance_analysis}


\subsection{Nebular Line Fluxes}
\label{ssec:nebular_line_fluxes}

\begin{table}
\caption{
Observed (F$_\lambda$) and de-reddened (I$_\lambda$; E(B--V)=0.46) nebular line intensities for SwSt~1 from
the 1993 UCLES spectrum. The de-reddened H$\beta$
flux is 1.50$\times$10$^{-10}$ erg~cm$^{-2}$~s$^{-1}$. Errors are gauged at 10\%.
The lines' rest wavelengths are from Kaufman and Sugar (1986).
The spectral resolution is indicated at the beginning of each Section.}

\begin{tabular}{@{}lllll@{}}
Ion & $\lambda_o$ & FWHM  &F$_\lambda$& 100$\times$I$_\lambda$ \\
    &  (\AA) & (km~s$^{-1}$)&(ergs~cm$^{-2}$~s$^{-1}$) &/I(H$\beta$)   \\
\multicolumn{5}{c}{$\Delta$v$\sim$900 km~s$^{-1}$ }\\
   C~{\sc ii}]  &2326   & -- & 3.9$\times$10$^{-12}$ & 97\\
\multicolumn{5}{c}{$\Delta$v$\sim$30 km~s$^{-1}$ }\\
   C~{\sc iii}] &1906.7 & -- & 5.7$\times$10$^{-13}$ & 13\\
   C~{\sc iii}] &1908.7 & -- & 2.0$\times$10$^{-12}$ & 38\\
\multicolumn{5}{c}{$\Delta$v$\sim$20 km~s$^{-1}$}\\
$[$O~{\sc ii}]  &3726.0 & 44 & 6.51$\times$10$^{-12}$ & 30.0\\
$[$O~{\sc ii}]  &3728.8 & 46 & 2.55$\times$10$^{-12}$ & 11.7\\
$[$Ne~{\sc iii}]&3868.8 & 40 & 3.35$\times$10$^{-14}$ & 0.48\\
$[$S~{\sc ii}]  &4068.6 & 41 & 5.23$\times$10$^{-13}$ &  2.2 \\
$[$S~{\sc ii}]  &4076.3 & 41 & 2.05$\times$10$^{-13}$ &  0.85\\
\multicolumn{5}{c}{$\Delta$v$\sim$8 km~s$^{-1}$} \\
H$\delta$       &4101.7 & 44 & 6.20$\times$10$^{-12}$ & 24.7\\
H$\gamma$       &4340.5 & 41 & 1.19$\times$10$^{-11}$ & 44.9\\
$[$O~{\sc iii}] &4363.2 & 31 & 7.11$\times$10$^{-14}$ &  0.27\\
He~{\sc i}      &4471.5 & 38 & 6.96$\times$10$^{-13}$ &  2.5\\
H$\beta$        &4861.3  & 41 & 3.22$\times$10$^{-11}$ &100\\
$[$O~{\sc iii}] &4958.9 & 34 & 3.80$\times$10$^{-12}$ & 11.4\\
$[$O~{\sc iii}] &5006.8 & 33 & 1.01$\times$10$^{-11}$ & 29.5\\
$[$Ar~{\sc iii}]&5191.8 & 33 & 1.38$\times$10$^{-14}$ & 0.04\\
$[$N~{\sc i}]   &5197.9 & 77 & 4.80$\times$10$^{-14}$ &  0.13\\
$[$N~{\sc i}]   &5200.2 & 73 & 2.91$\times$10$^{-14}$ &  0.08\\
\multicolumn{5}{c}{$\Delta$v$\sim$20 km~s$^{-1}$}\\
$[$O~{\sc i}]   &5577.3 & 30 & 7.83$\times$10$^{-14}$ &  0.19\\
$[$N~{\sc ii}]  &5754.6 & 28 & 1.42$\times$10$^{-12}$ &  3.3\\
He~{\sc i}      &5875.7 & 42 & 3.17$\times$10$^{-12}$ &  7.1\\
$[$O~{\sc i}]   &6300.3 & 49 & 7.43$\times$10$^{-13}$ &  1.5\\
$[$S~{\sc iii}] &6312.1 & 39 & 1.30$\times$10$^{-12}$ &  2.6\\
$[$O~{\sc i}]   &6363 & 49 & 2.45$\times$10$^{-13}$ &  0.49\\
$[$N~{\sc ii}]  &6548.0 & 46 & 1.64$\times$10$^{-11}$ & 31.1\\
He~{\sc i}      &6678.1 & 39 & 9.26$\times$10$^{-13}$ &  1.7\\
$[$S~{\sc ii}]  &6716.5 & 62 & 3.91$\times$10$^{-13}$ &  0.72\\
$[$S~{\sc ii}]  &6730.8 & 55 & 7.92$\times$10$^{-13}$ &  1.4\\
He~{\sc i}      &7065.2 & 39 & 2.11$\times$10$^{-12}$ &  3.6\\
$[$O~{\sc ii}]  &7318.6 & -- & 3.96$\times$10$^{-12}$: &  6.4\\
$[$O~{\sc ii}]  &7319.4 & -- & 4.35$\times$10$^{-11}$: & 70.7\\
$[$O~{\sc ii}]  &7329.6 & 33 & 8.36$\times$10$^{-12}$ & 13.6\\
$[$O~{\sc ii}]  &7330.7 & 35 & 8.62$\times$10$^{-12}$ & 14.0\\
$[$Ar~{\sc iii}]&7751.1 & 34 & 7.13$\times$10$^{-13}$ & 1.1 \\
$[$S~{\sc iii}] &9531.0 & 38 & 7.53$\times$10$^{-11}$ & 90.7\\
\label{tab:nebular_fluxes}
\end{tabular}
\end{table}

The nebular spectrum of SwSt~1 is much stronger and slightly higher in
excitation than that of the PN with [WC10] central stars analyzed by DBS97
and De Marco \& Crowther (1998).  Many nebular He~{\sc i} lines could be
measured: amongst them we selected those at 4471.5, 5875.7 and 6678.1 \AA
, disregarding the line at 7065.2 \AA\ due to its being dominated by
collisions and the possibility that it is affected by telluric
absorption. The [O~{\sc ii}] line complex at 7325 \AA\ is made up of 4
lines with wavelengths 7318.8, 7319.9, 7329.6, and 7330.6~\AA , where the
individual components are partly blended. Unfortunately, the blend at 7319
\AA\ was affected by an echelle order join which removed a part of the
flux. In order to obtain a measurement of the flux in those lines we
therefore snipped the section containing the join. The snipped line,
despite missing the central part, still had clearly identifiable wings and
could therefore be successfully fitted with a Gaussian. We nonetheless
decided to assign a 50\% uncertainty to it and did not use it in any of
the parameter determinations. The [O~{\sc ii}] lines at 3726.0 and 3728.8
\AA\ are resolved and completely un-blended with any stellar feature.
[O~{\sc iii}] lines appear at 4363.2, 4958.9 and 5006.8~\AA .

All four [S~{\sc ii}] nebular lines included
in our wavelength range could be measured, namely the lines at 4068.6, 4076.3, 6716.5
and 6730.8 \AA . The line at 4068.6 \AA\ was found to lie on an order join. However
since abundance results obtained with it or with the weaker lines at 4076.3 \AA\ were
very similar, we decided to
treat it normally and only assign it a marginally higher uncertainty.
To determine the abundance of S$^{2+}$, we used $\lambda$9531.0.
The telluric water vapour line at 9069.126 \AA\ affects the measurement of
the
[S~{\sc iii}] line at 9068.9 \AA\ since its measured ratio to the
$\lambda$9531 line 
is 0.28, lower than the theoretical ratio of 0.403.

Three lines of singly ionized nitrogen reside in our spectral range, but as the
[N~{\sc ii}] ratio
$\lambda$6583/$\lambda$6548 was equal to 1.4,
instead of 2.90, we decided that the strong $\lambda$6583 line was
saturated (as was H$\alpha$) and did not include it in any of our determinations.
Together with the $\lambda$6548 line we used [N~{\sc ii}] $\lambda$5755.

\begin{figure}
\vspace {6cm}
\caption{The high resolution IUE spectrum of SwSt~1 showing the semi-forbidden
stellar-nebular blend of C~{\sc iii}] $\lambda$1906,1909.} 
\includegraphics{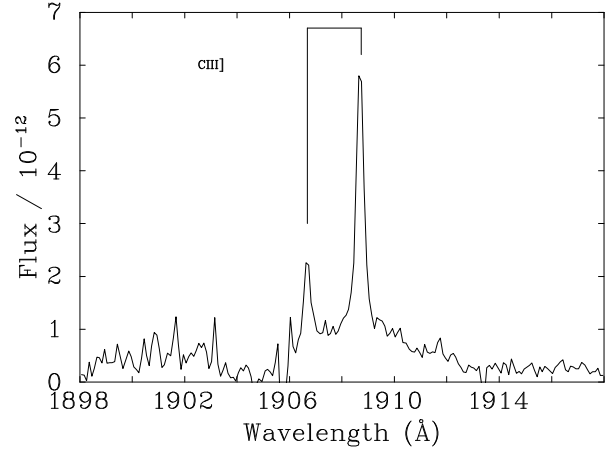}
\label{fig:iue}
\end{figure}

To derive the nebular C/H ratio we used IUE observations of the
semi-forbidden 
C~{\sc ii}] $\lambda$2326 and C~{\sc iii}] $\lambda$1909 lines.
We assume the stellar contribution to $\lambda$2326
to be negligible, while we could de-blend the stellar and nebular
contributions to
C~{\sc iii} $\lambda$$\lambda$1906,1909 using the high resolution IUE spectrum
(Fig.~\ref{fig:iue}).

Besides the lines listed in Table~\ref{tab:nebular_fluxes}, we identify
forbidden lines
of Fe$^{+}$ (e.g. [Fe~{\sc ii}] $\lambda$4606), Fe$^{2+}$ (e.g. [Fe~{\sc iii}] $\lambda$4881)
and possibly Fe$^{3+}$ (e.g. [Fe~{\sc iv}] $\lambda$5677). We also identify
[Ne~{\sc iii}] $\lambda$3868, [Ar~{\sc iii}] $\lambda$5191.8 $\lambda$7135.8 (affected by an order join)
and $\lambda$7751.1 and
[Cl~{\sc iii}] $\lambda$5517.7 and $\lambda$5537.9. A line list is presented in 
Appendix~B.

\subsection{Nebular Temperature, Density and Abundances}
\label{ssec:nebular_temperature_density_and_abundances}

\begin{table}
\caption
{The nebular electron temperature and density adopted for SwSt~1, along with
values from the literature.}
\renewcommand\arraystretch{0.7}
\begin{tabular}{lccc}
T$_e$ (K) & log(N$_e$) & E(B--V)& Ref.\\
10\,500 $\pm$ 500 & 4.5 $\pm$ 0.2 & 0.46 & this work \\
11\,400 $\pm$ 500 & 5.0 $\pm$ 0.1 & 0.41 & dFPV87 \\
8000             & 5.0 $\pm$ 0.1 & 0.34 & FGC84    \\
\end{tabular}
\label{tab:ne_te}
\end{table}

\begin{figure}
\vspace {6cm}
\includegraphics{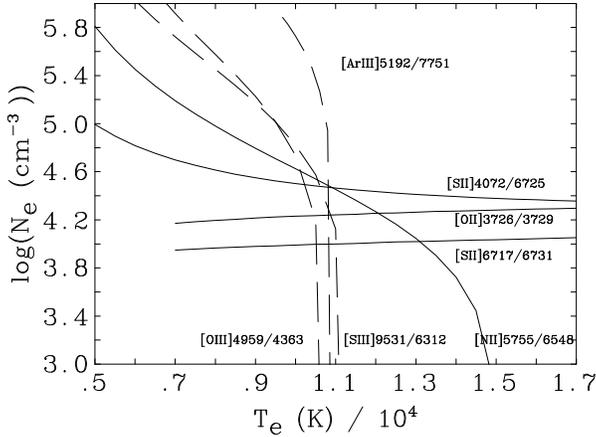}
\caption{Diagnostic diagram,
log(N$_e$) vs T$_e$, for SwSt~1. The solid lines are obtained from ratios of ions
in the first stage of ionization, the dashed lines are for ions in the second
ionization stage.}
\label{fig:ne_te}
\end{figure}

In Fig.~\ref{fig:ne_te} we plot a nebular diagnostic diagram for SwSt~1. From a weighted mean
of the diagnostic lines we adopt values for the
electron temperature and density of 10\,500 K and 31\,600 cm$^{-3}$ (log(N$_e$)=4.5),
respectively (Table~\ref{tab:ne_te}).

The temperature implied by the [O~{\sc iii}], [S~{\sc iii}] and 
[Ar~{\sc iii}] diagnostic ratios is between 11000~K and 10000~K for
electron densities between 10$^3$ and 10$^5$~cm$^{-3}$. 
The C~{\sc iii}] $\lambda$1909/$\lambda$1906 ratio of 0.34 indicates
$\log$$N_e$=5.1 for $T_e$ in the range 9000--14000~K. 
However this line ratio is severely affected by the blend with the stellar
component. De-blending by fitting nebular and stellar components to the
doublet resulted in a large uncertainty.

The diagnostic ratios from singly ionized elements seem to agree on a lower density (4.0$<$log(N$_e$)$<$4.5), if
we make an exception for the [O~{\sc ii}] ratio $\lambda$3727/$\lambda$7330.
This ratio uses lines from
two different grating settings and therefore might be affected by errors in the flux calibration.
However if this were the cause of the discrepancy we would also have to be wary of the
[S~{\sc iii}] $\lambda$9531/$\lambda$6312 ratio (which we have used above) and the
[S~{\sc ii}] $\lambda$4072/$\lambda$6725 ratios, although these employ
the green and red settings and not the blue and red settings used by the
[O~{\sc ii}] $\lambda$3727/$\lambda$7330 ratio.
The ratios [O~{\sc ii}] $\lambda$3727/3729 and [S~{\sc ii}]
$\lambda$6717/6731 are unreliable
because they are at their high density limit. Therefore only [S~{\sc ii}]
$\lambda$4072/$\lambda$6725
yields a reliable density diagnostic.
From [S~{\sc iii}] $\lambda$6312/$\lambda$9531, [O~{\sc iii}]
$\lambda$4959/$\lambda$4363, [Ar~{\sc iii}] $\lambda$7751/$\lambda$5192,
[N~{\sc ii}] $\lambda$5755/$\lambda$6548 and
[S~{\sc ii}] $\lambda$4072/$\lambda$6725 we derive T$_e$=10\,500~K and
log(N$_e$/cm$^{-1}$)=4.5.

Although the HST images point to a somewhat stratified PN (with the [O~{\sc iii}] image being
smaller than the H$\beta$ one) we feel inclined to average the results obtained from the
singly and doubly ionized elements.
SwSt~1 is a young and compact PN and we therefore feel unjustified in using different electron
density and temperature combinations for the singly and doubly ionized regions, a choice
that is more appropriate for more extended, higher ionization objects. 

Flower et al. (1984) also
derived log(N$_e$)=5.0 for SwSt~1 from the high resolution IUE spectrum of the
C~{\sc iii}] $\lambda$$\lambda$1907,1909 diagnostic lines.
They also quoted
an upper limit for the electron density of 2$\times$10$^4$ cm$^{-3}$, derived from
the [O~{\sc ii}] $\lambda$3729/$\lambda$3726 ratio
(which is almost identical to ours), however they argued that this ratio is too close
to the high density limit
to be reliable. 
Their temperature determination, using the [O~{\sc iii}] $\lambda$5007/$\lambda$4363 ratio, was
8800 K (for E(B--V)=0.70), not inconsistent with our own value (although
we employ a lower reddening, the effect is not major over such a short baseline). However
their determination using the [O~{\sc iii}] $\lambda$5007/$\lambda$1663 ratio
implies a higher electron temperature (15\,000~K). If we ascribe the discrepancy to the
use of two different spectra in the second determination (IUE and optical), which might have imperfect absolute
flux calibrations, we can reconcile
our own determinations of the temperature and density for the doubly ionised element with
theirs.
de~Freitas Pacheco and Veliz (1987) derived log(N$_e$/cm$^{-3}$)=5.0 and 
T$_e$=(11\,400 $\pm$ 500)~K, from
[S~{\sc ii}] $\lambda$$\lambda$4068,76/$\lambda$$\lambda$6716,31,
[O~{\sc iii}] $\lambda$4363/($\lambda$4959+$\lambda$5007) and
[N~{\sc ii}] $\lambda$5755/$\lambda$$\lambda$6548,84. Their
diagnostic diagram is not inconsistent with our own.

\begin{table}
\caption {Nebular ionic abundances for SwSt~1.} 
\begin{tabular}{lll}
Ratio             &   Abundance & Lines \\
                  &              &Used \\
He$^+$/H$^+$      & 0.044    $\pm$ 0.004    & He~{\sc i} $\lambda$$\lambda$4471,5876,\\
C$^{+}$/H$^{+}$   & 1.21(-4)$^a$ $\pm$ 0.60(-4) & C~{\sc ii}]$\lambda$2326\\
C$^{2+}$/H$^{+}$  & 1.19(-4) $\pm$ 0.24(-4) & C~{\sc iii}]$\lambda$1909\\
N$^{+}$/H$^{+}$   & 1.99(-5) $\pm$ 0.20(-5) & $[$S~{\sc ii}]$\lambda$$\lambda$5755,6548\\
O$^{+}$/H$^{+}$   & 2.44(-4) $\pm$ 0.24(-4) & $[$O~{\sc ii}]$\lambda$$\lambda$3726,3729,7330\\
O$^{2+}$/H$^{+}$  & 1.05(-5) $\pm$ 0.10(-5) &$[$O~{\sc iii}]$\lambda$4959\\
Ne$^+$/H$^{+}$    & 9.77(-5) $\pm$ 1.19(-5) & [Ne~{\sc ii}] 12.8 $\mu$m$^b$\\
Ne$^{2+}$/H$^{+}$ & 3.94(-6) $\pm$ 0.80(-6) & [Ne~{\sc iii}] $\lambda$3868\\
S$^{+}$/H$^{+}$   & 2.96(-7) $\pm$ 0.30(-7) & $[$S~{\sc ii}]$\lambda$$\lambda$6713,6731,\\
S$^{2+}$/H$^{+}$  & 4.23(-6) $\pm$ 0.80(-6) & $[$S~{\sc iii}]$\lambda$9531\\
\noalign{\smallskip}
\multicolumn{3}{l}{$^a$1.21(-4)=1.21$\times$10$^{-4}$}\\
\multicolumn{3}{l}{$^b$line flux from Aitken and Roche 1982\nocite{AR82}}\\
\end{tabular}
\label{tab:nebular_abundances_ionic}
\end{table}

Our derived abundances are listed in Tables~\ref{tab:nebular_abundances_ionic}
and \ref{tab:nebular_abundances_total}, where they are compared with the
mean values for
Type-I and non Type-I PN from Kingsburgh \& Barlow (1994). The quoted errors derive from
errors on the flux measurements and do not take into account errors on the electron temperature
and density.
The C/H number ratio of 2.4$\times$10$^{-4}$ is not as high as for CPD-56$^{\rm o}$8032 
(63$\times$10$^{-4}$;
DBS97), He~2-113 (50$\times$10$^{-4}$; DBS97) or M~4-18 (12$\times$10$^{-4}$; 
De Marco \& Crowther 1998) and lower even than the
mean PN value as determined by Kingsburgh and Barlow (1994; 5.49$\times$10$^{-4}$). 
Additionally, with C/O=0.94, SwSt~1 qualifies as an O-rich PN
Flower et al. (1984) obtained C/O=0.72, while de Freitas Pacheco and Veliz
(1987) derived C/O=0.54. The C/O ratio is consistent with the presence of
a silicate emission feature at 10~$\mu$m (see Aitken \& Roche 1982 and
the SWS spectrum plotted in Fig.~\ref{fig:model_flux}).

To obtain the N/H number ratio we multiplied the N$^+$/H$^+$ ratio by an ionization correction
factor of 1.05, obtained from the ratio of total oxygen to singly ionized oxygen abundances,
in accordance with Kingsburgh and Barlow (1994).

\section{HST Images}
\label{sec:hst_images}

\begin{figure*}
\vspace{7cm}
\includegraphics{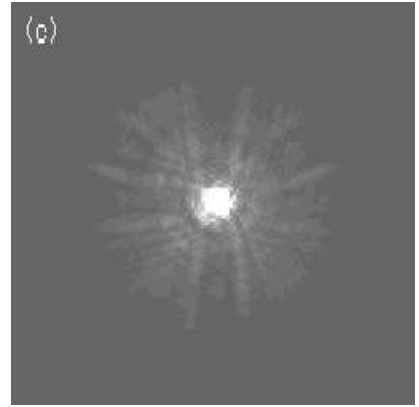}
\caption[HST images of SwSt~1]{HST images of SwSt~1;
a) The average of the two raw H$\beta$ images; b) the average of the two
raw [O~{\sc iii}] images; c) the synthetic point spread function generated
by the program Tiny Tim; the FOV is 6~arcsec, north is towards the top, east to the left.}
\label{fig:hst}
\end{figure*}

Perek and Kohoutek (1967) quote an optical angular diameter of 5~arcsec,
while Aaquist and Kwok (1990) measured a 5~GHz radio diameter of only 1.3~arcsec.
No resolved optical image of SwSt~1 was found in the literature. 
SwSt~1 was observed in 1992 by HST in the H$\beta$ and 
[O~{\sc iii}] $\lambda$5007 lines, as part of
a WF/PC snap--shot survey (two exposures per filter). 
These four pre-COSTAR images were already calibrated by the pipeline process
(details can be found in the HST data handbook).
DBS97 used the stellar [O~{\sc iii}] images as a point spread function to 
de-convolve their images of CPD--56$^{\rm o}$8032 and He~2--113.
However the nebula around SwSt~1 is more highly ionised than
the ones around CPD--56$^{\rm o}$8032 and He~2--113, 
as revealed by the prominent [O~{\sc iii}]
lines observed in our spectrum. Hence the [O~{\sc iii}] image could not be used
as a point spread function to de-convolve the H${\beta}$ image.
We therefore used the program Tiny Tim (V4.0\footnote{Tiny Tim is supported by 
the Space Telescope Science Institute;  
http://www.stsci.edu/software/tinytim/.})
to create theoretical PSFs for the H$\beta$ and [O~{\sc iii}] images
(the theoretical H$\beta$ PSF is shown in Fig.~\ref{fig:hst} together with the
raw images).

\begin{figure*}
\vspace{3.2in}
\caption{De-convolved H$\beta$ (a) and [O~{\sc iii}] (b) HST images of SwSt~1;
The images are displayed on a log scale; 
the FOV is 2.6~arcsec, north is towards the top, east to the left.}
\label{fig:hst2}
\end{figure*}

The de-convolved images of SwSt~1 (Fig.~\ref{fig:hst2})
reveal a broken ring of diameter ((1.3$\times$0.9) $\pm$ 0.2)~arcsec, elongated in the
east--west direction, in agreement with the radio images of Kwok et
al. (1981) and
Aaquist and Kwok (1990), but smaller than the 5~arcsec quoted by Perek and Kohoutek (1967).
Although the images are of poor quality, this structure
could be a ring whose plane is at about 30$^o$ from the line of sight.
If we take 1.3~arcsec as the true diameter, a distance of 2.0~kpc and the expansion velocity of
21~km~s$^{-1}$ would imply a dynamical time-scale of 290~yr 
for the outer edge of the ionized
material (although we do not know how far the neutral shell extends).

The [O~{\sc iii}] image appears smaller than the H$\beta$ image (about 75\% of the area),
as can be seen from overlaying the normalised contours of the [O~{\sc
iii}] image onto the
grey--scale plot of the H~$\beta$ image (Fig.~\ref{fig:hst3}). This is consistent
with ionization stratification of the PN.
Further confirmation of the stratified structure of this PN comes from analysing the
nebular line widths as a function of ionic charge (Fig.~\ref{fig:line_shapes}).
After de-convolving the FWHM of individual lines with the respective instrumental profiles,
nebular forbidden lines arising from singly ionised ions
(e.g. [N~{\sc ii}] $\lambda$6548 and [S~{\sc ii}] $\lambda$6717), as well as lines coming
from neutral ions (e.g. [O~{\sc i}] $\lambda$6300) are systematically broader
(FWHM$\sim$45--60 km~s$^{-1}$), while lines arising from doubly ionised
ions (e.g. [O~{\sc iii}] $\lambda$5007 and [Ar~{\sc iii}] $\lambda$7750) 
have widths similar to the Balmer lines (FWHM(H$\beta$)=40.2 km~s$^{-1}$)
or narrower (FWHM$\sim$30--40 km~s$^{-1}$; see Fig.~\ref{fig:line_shapes}).
This is consistent with the
lines arising from higher ionization stages being formed in an inner, slower part of
the PN, and the lines from lower ionization stages being formed further out
in an accelerating part of the flow.
The Balmer emission, coming from the entire ionised region,
might have been expected to produce broader lines,
although it might be more intense in the
inner, denser region of the PN, yielding narrower profiles.

\begin{figure}
\vspace{8cm}
\caption{De-convolved H$\beta$ HST image of SwSt~1. Over-plotted are the 
contours from the [O~{\sc iii}] image (Fig.~\ref{fig:hst2}).}
\label{fig:hst3}
\end{figure}

\begin{figure}
\vspace{8cm}
\includegraphics{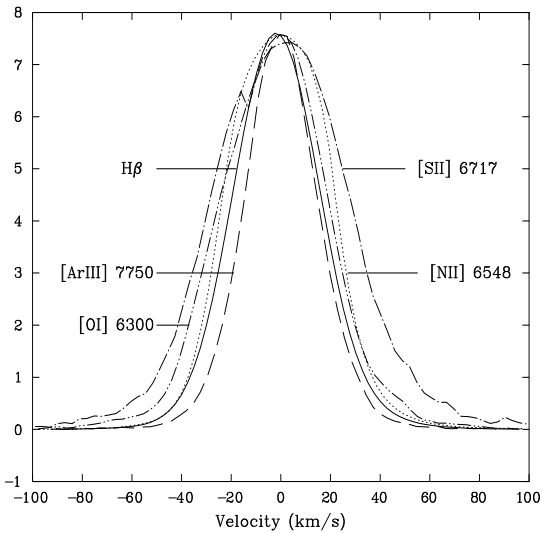}
\caption
{Comparison of selected nebular lines in the AAT--UCLES spectrum of
SwSt~1. The lines
are background subtracted and normalised to the peak intensity of the
[Ar~{\sc iii}] $\lambda$7750 line.}
\label{fig:line_shapes}
\end{figure}

\section{Nebular Modelling}
\label{sec:nebular_modelling}

\subsection{Zanstra Temperatures}
\label{ssec:zanstra_temperatures}

Following the formula of Milne and Aller (1975), for a 5~GHz flux of 
243 mJy (the weighted average of
Kwok et al. 1981, Milne and Aller 1982 and Aaquist and Kwok 1990, see 
Section~\ref{ssec:reddening}),
we predict an intrinsic H$\beta$ flux of 
8.07$\times$10$^{-11}$ ergs~cm$^{-2}$~s$^{-1}$. This H$\beta$ flux
agrees within the uncertainties
with the fluxes of Flower et al.
(1984) for their adopted reddening of c(H$\beta$)=0.68.
From these de-reddened H$\beta$ fluxes and the distance of 2.0 kpc
for SwSt~1 we derive a flux of hydrogen--ionizing photons for the star
corresponding to 
log(Q$_o$(s$^{-1}$))=46.90.

We carried out a Zanstra analysis using blackbody as well
as  Kurucz ATLAS9 (Kurucz 1991)\nocite{K91} model atmospheres
with log(g)=4.0, the minimum gravity tabulated for atmospheres of the required effective temperatures.
We derive a blackbody H~{\sc i} Zanstra temperature
of (30\,300 $\pm$ 500)~K,
while the ATLAS9 model grid yields an
effective temperature of (34\,000 $\pm$ 500)~K.
The radius predicted for the scaled blackbody is 2.7 R$_\odot$ at the
adopted distance of 2.0~kpc, from which a luminosity of 3550~L$_\odot$ can be derived.
A similar exercise for the Kurucz atmospheres yields a radius of 2.3 R$_\odot$ 
and a stellar luminosity of 6300~L$_\odot$.

de Freitas Pacheco and Veliz (1987)
derived a blackbody H~{\sc i} Zanstra temperature of 32\,000 K.
A comparison  by Flower et al. (1987)
of the stellar ultraviolet and nebular free--free radio continuum fluxes yielded a
blackbody H~{\sc i} Zanstra temperature
of 36\,000 K, although they found that a 30\,000 K blackbody
better fitted their optical and UV de-reddened stellar energy distribution.

\begin{table*}
\begin{minipage}{110mm}
\caption
{Empirical and modelled nebular abundances for SwSt~1 compared with mean values for Type-I and
non Type-I PN from Kingsburgh \& Barlow (1994).}
\begin{tabular}{lllll}
Ratio             & Abundance          & Abundance & Abundance & Abundance \\
                  & Empirical          & Model$^b$ & Type-I    & non Type-I \\
He$^+$/H$^+$      & 0.044 $\pm$0.004   &0.04       & 0.129 $\pm$ 0.037 & 0.112 $\pm$ 0.015\\
$\log$(C/H)+12    & 8.38 $\pm$ 0.25    &8.40       & 8.48  $\pm$ 0.30  & 8.81 $\pm$ 0.30 \\
$\log(N/H)$+12    & 7.32$^a$ $\pm$ 0.10&7.99       & 8.72  $\pm$ 0.15  & 8.14 $\pm$ 0.20\\
$\log$(O/H)+12    & 8.41 $\pm$ 0.20    &8.15       & 8.65  $\pm$ 0.15  & 8.69 $\pm$ 0.15 \\
$\log$(Ne/H)+12   & 8.01 $\pm$ 0.07    &7.88       & 8.09  $\pm$ 0.15  & 8.10 $\pm$ 0.15\\
$\log$(S/H)+12    & 6.65 $\pm$ 0.10    &6.62       & 6.91  $\pm$ 0.30  & 6.91 $\pm$ 0.30\\
C/O               & 0.94               &1.79       & 0.68  & 1.32\\
N/O               & 0.08               &0.08       & 1.17  & 0.28\\
\noalign{\smallskip}
\multicolumn{5}{l}{$^a$ICF(N)=O/O$^+$=1.05, N/H=IFC(N)$\times$N$^+$/H$^+$, following Kingsburgh \& Barlow 1994}\\
\multicolumn{5}{l}{$^b$Abundances derived from the photoionization model, using the {\sc isa}-Wind WR }\\
\multicolumn{5}{l}{    stellar atmosphere as input (Table~\ref{tab:nebular_model}).}\\
\end{tabular}
\label{tab:nebular_abundances_total}
\end{minipage}
\end{table*}

\subsection{Modelling Strategy and Results}
\label{ssec:modelling_strategy}

Photo-ionization modelling of PN is the best way to test the internal
consistency of an empirical analysis.
The Harrington photo-ionization code applied here (Harrington et al. 1982) assumes that
the nebula can be represented by a hollow spherical shell which is ionised
only by a central star, and sampled by 60 radial grid points. 

The Kurucz model and blackbody luminosities and the distance 
were fixed at the values derived above.
An {\it inner} nebular angular radius of 0.22~arcsec (approximately 5 pixels)
was determined from the HST observations, and scaled to the adopted distance (2.0~kpc) to
yield an inner radius of 0.0086~pc.
The clumping and vacuum filling factors were varied so as to reproduce the de-reddened H$\beta$ flux.
The electron density was kept constant throughout the PN, fixed at the empirically derived value,
except in the case of the WR flux distribution, for which this value overestimated the H$\beta$ flux
even for extremely low values of the filling factor, a fact that was
judged non-physical.
In theory, such a high density PN is likely to be optically thick, such that a correct modelling procedure
should make sure that the model reaches far enough to include the edge of the Str\"{o}mgren sphere.
This is achieved by increasing the thickness of the PN shell until the the PN recombines at the outer grid points.
On the other hand, following this procedure would mean that 
the thickness of the ionized shell and the corresponding H$\beta$ flux would be much larger than observed, a fact that
cannot be compensated for by reducing the filling factor within sensible limits.
We therefore used the observed values for the inner and outer radii and accepted an optically thin PN.
A non-constant density profile was not attempted. The efforts of De Marco \& Crwother (1999) on the
PN M4-18 did not produce better results when a variable density profile was adopted. 
Additionally, for the compact PN SwSt~1 the lack of a high signal-to-noise ratio image
means that the density profile cannot be constrained.

As a starting point we adopted the empirically derived abundances. Lines
of oxygen and carbon were then fitted by changing their abundances. Carbon
and oxygen are important coolants and changing the abundances not
only varies the line strengths, but also the nebular electron temperature.
When a compromise was reached on the strength of these lines and the
electron temperature, lines of sulphur, nitrogen and neon were fitted by
varying their abundances. The abundances of other unobserved species were
left at the
mean PN value as determined by Kingsburgh and Barlow (1994) for a set of
80 southern PN, or at their solar values. When the ionization balance of
the element in question is very different from the observations it is
arbitrary to decide which lines should be fitted and which should be
ignored. In the case of sulphur, we used the strong [S~{\sc iii}] line at
9532~\AA , ignoring [S~{\sc ii}] lines. For oxygen, we fitted a compromise
of the [O~{\sc ii}] lines $\lambda$$\lambda$3726,29 and $\lambda$7325,
forsaking [O~{\sc iii}] lines. For neon, the [Ne~{\sc ii}] line at
12.8~$\mu$m was used, while for carbon a compromise between the 
C~{\sc ii}] $\lambda$2326 and C~{\sc iii}] $\lambda$1908 lines was
used. We note however
that because none of the stellar flux distributions reproduces the
observations to an acceptable precision, we found it unnecessary to fit
individual spectral line fluxes to high accuracy.

\begin{table}
\caption{Comparison of results from photo-ionization models for
SwSt~1 with observations (I(H$\beta$)=100),
using {\sc isa}-Wind non-LTE, blackbody and Kurucz plane--parallel energy distributions.}
\label{tab:nebular_model}
\begin{tabular}{lccccc}
Parameter              &Observed& WR   &Blackbody&Kurucz\\
$T_{\em eff}$(K)       & --     & 40.0  & 30.3  & 34.0      \\
log($L/L_{\odot})$     & --     & 3.94  & 3.74  & 3.80    \\
R(R$_\odot$)           & --     & 2.0   & 2.7   & 2.3      \\
log $Q_{0}$(s$^{-1}$)  & 46.90  & 47.53 & 46.90 & 46.90     \\
log(I(H$\beta$))       & --10.1 & --10.1& --10.1& --10.1  \\
N$_e$(cm$^{-3})$       & 31620  &15000  & 31620 & 31620 \\
$\epsilon$             & --     &0.76   & 0.15  & 0.18    \\
R$_{inner}$(pc)        & 0.0086 &0.0086 & 0.0086& 0.0086 \\
R$_{outer}$(pc)        & 0.0252 &0.0252 & 0.0252& 0.0252 \\
M$_{PN}$(M$_\odot$)    & --     &0.021  & 0.009 & 0.015  \\
$\tau$(H~{\sc i})      & --     &0.69   & 1.3   & 2.2     \\
\noalign{\smallskip}
He/H                      &0.04 &0.04 & 0.064 & 0.15  \\
C/H$\times$10$^4$         &2.51 &2.51 & 4.68  & 2.51   \\
N/H$\times$10$^5$         &1.99 &9.86 & 4.11  & 2.13  \\
O/H$\times$10$^4$         &2.57 &1.40 & 1.90  & 0.78  \\
S/H$\times$10$^6$         &4.17 &4.00 & 6.00  & 4.60  \\
Ne/H$\times$10$^5$        &9.80 &7.60 & 8.80  & 7.90  \\
\noalign{\smallskip}
1909  C\,{\sc iii}]    & 71     & 68  & 33     & 14   \\
2326  C\,{\sc ii}]     & 85     & 19  & 76     & 208  \\
3726 [O\,{\sc ii}]     & 30     & 57  & 39     & 38   \\
3729 [O\,{\sc ii}]     & 18     & 23  & 14     & 14   \\
3868 [Ne~{\sc iii}]    & 0.5    & 0.0 & 4.8    & 0.06\\
4068 [S\,{\sc ii}]     & 2.2    & 2.5 & 2.5    & 3.7  \\
4076 [S\,{\sc ii}]     & 0.85   & 0.27& 0.82   & 1.2  \\
4363 [O\,{\sc iii}]    & 0.3    & 0.09& 0.4    & 0.02\\
4471 He\,{\sc i}       & 2.5    & 2.3 & 3.3    & 1.2  \\
5007 [O\,{\sc iii}]    & 29     & 204 & 62     & 2.2  \\
5755 [N\,{\sc ii}]     & 3.2    & 2.2 & 1.9    & 3.1  \\
5876 He\,{\sc i}       & 7.1    & 6.7 & 9.2    & 3.6  \\
6300 [O\,{\sc i}]      & 1.5    & 0.0 & 0.06   & 0.07 \\
6363 [O\,{\sc i}]      & 0.5    & 0.0 & 0.02   & 0.02\\
6548 [N\,{\sc ii}]     & 31     & 30  & 27     & 32   \\
6678 He\,{\sc i}       & 1.7    & 1.8 & 2.5    & 1.0  \\
6717 [S\,{\sc ii}]     & 0.7    & 0.2 & 0.8    & 1.1  \\
6731 [S\,{\sc ii}]     & 1.4    & 0.4 & 1.9    & 2.5  \\
7325 [O\,{\sc ii}]     & 28     & 18  & 20     & 23   \\
9532 [S~{\sc iii}]     & 91     & 84  & 92     & 94  \\
128\,000 [Ne\,{\sc ii}]& 68     & 67  & 65     & 67   \\
\smallskip
C$^{2+}$/C$^{+}$       & 0.98   & 9.8    & 0.23   & 0.17 \\
O$^{2+}$/O$^{+}$       & 0.043  & 0.81   & 0.18   & 0.088\\
S$^{2+}$/S$^{+}$       & 14     & 9.7    & 10.7   & 15   \\
Ne$^{2+}$/Ne$^{+}$     & 0.040  & --     & 0.063  & 0.004\\
$T_{e}$(N$^+$)(K)      & 10\,500&10\,500   &9300  & 10\,800 \\

\end{tabular}
\label{tab:nebular_model}
\end{table}

In Table~\ref{tab:nebular_model} we compare the abundances determined from
modelling with the empirical values, as well as the predicted line fluxes
with the de-reddened observations. None of the stellar atmospheres used
reproduces the observed characteristics of SwSt~1's PN to an acceptable
degree of accuracy. In the case of the WR flux distribution, we decided
that lowering the electron density to about half the observed value
was preferable to having an extremely low filling factor ($\sim$0.02, too low
even if the PN is actually a ring as the HST image might suggest).
With this value of the electron density, a more acceptable filling factor
of 0.76 was needed to reproduce the H$\beta$ flux. However, this implies
an
optically thin PN. The overall ionization balance is overestimated except
that of sulphur, which is underestimated. Interestingly, a cooler WR
stellar atmosphere, also failed to reproduce the ionization balance. We
tried to use the 35kK model which reproduces most of the stellar spectral
features except the C~{\sc iv}/C{\sc iii} ionization balance (see
Sect.~\ref{ssec:stellar_model_results}). This model severely
underestimates the nebular ionization balance, while at the same time
having a
similar problem with the H$\beta$ flux and the filling factor, as outlined
above. We therefore believe that the failure of the PN model is
not due to the hot WR atmosphere.

\begin{figure*}
\vspace{7.5cm}
\includegraphics{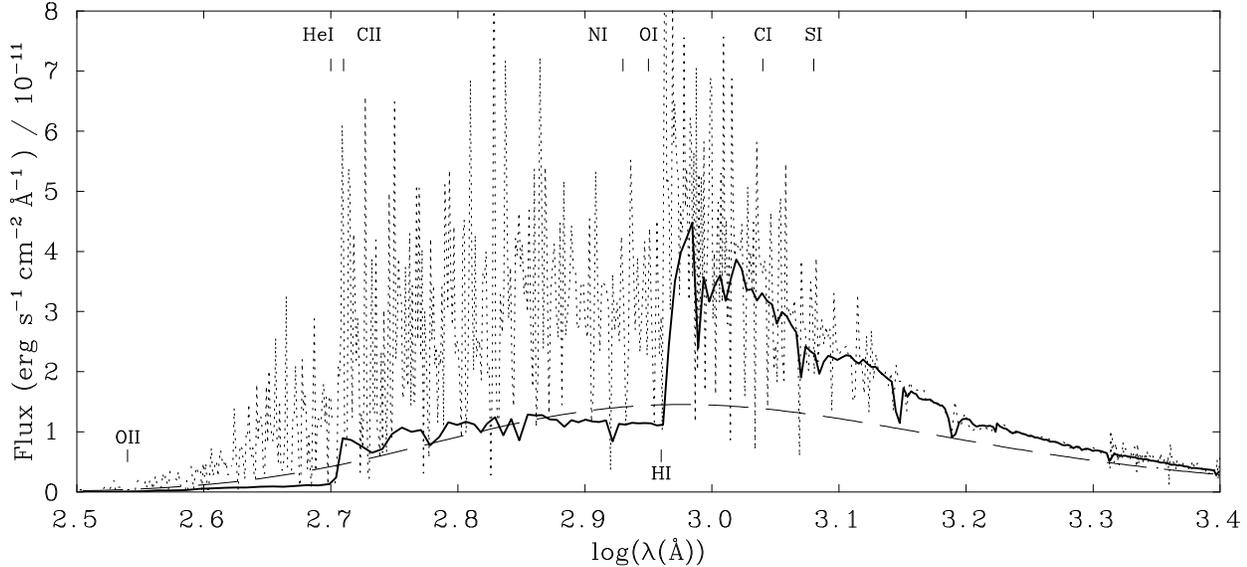}
\caption
{A comparison of the {\sc isa}-Wind line-blanketed non-LTE model atmosphere 
(dotted line) with the 
30\,300 K blackbody flux distribution (dashed line),
and a 34\,000 K, log(g)=4.0 plane--parallel
Kurucz model atmosphere (thick solid line). The H~{\sc i} and He~{\sc i} and other important ionization edges are marked.}
\label{fig:zanstra}
\end{figure*}

The blackbody and Kurucz stellar flux distributions underestimate the
ionization balance of carbon, but overestimate that of oxygen. They give
mixed results in the case of neon and sulphur. With particular reference to
oxygen, we point out that the [O~{\sc iii}] lines used are extremely susceptible to
small temperature changes, in particular it appears that for central star 
effective temperatures above $\sim$38kK the [O~{\sc iii}] lines increase 
dramatically in flux (Miriam Pe\~na, priv. comm.). For these stellar flux
distributions we were able to use the electron density determined from
observations, although, once again, we had to accept an optically thin
PN, so as to obtain a sensible value for the filling factor. An optically
thin PN is however more in disagreement with an electron density of
31\,600~cm$^{-3}$. The three stellar atmospheres are compared in
Fig.~\ref{fig:zanstra}.

The nebular mass implied by all the stellar flux distributions is very low
compared to the canonical value of 0.3~M$_\odot$ for optically thin
PN. As for the optical depth
properties of this PN, it is unusual to have an optically thin PN with an
electron density as high as we encountered. On the other hand the fact
that the Zanstra analysis returns values of the effective temperature
lower than determined with an {\it ad hoc} modelling of the stellar
spectrum with a non-LTE code, might suggest that the H$\beta$ flux is not
representative of the hard photon population, i.e. that the PN is indeed
optically thin.

In conclusion, no matter the method chosen for the
modelling effort, we have found no way of fitting the observed properties
of
this PN to a satisfactory accuracy. We take this to mean that the
modelling tool being used is inadequate for the PN of SwSt~1. In particular
the simple shell geometry might be far from the truth of this young PN.
Nebular modelling with our {\sc cmfgen} models atmosphere as input was not attempted.
Overall the two flux distributions compare reasonably well, with {\sc isa}-Wind 
harder far UV flux being compensated by {\sc cmfgen}'s higher effective temperature.  
{\sc cmfgen} would therefore not succeed in reproducing the [O{\sc iii}] lines.

\section{SwSt~1 between 1895 and 1993}
\label{sec:swst1_between_1895_and_1993}

The warm oxygen-rich dust revealed by the 10~$\mu$m silicate emission
feature, and the high nebular density, indicate that SwSt~1 has recently
made the transition from the AGB. This could indicate that the pulse that
caused SwSt~1 to become H-deficient might have occurred in the last
century, i.e. after it was first observed in 1895. Since variability of
the stellar and nebular lines has been noted by Carlson \& Henize (1997)
and by de~Freitas Pacheco \& Veliz (1987), we decided to investigate the
variability issue.
Photographic spectra of SwSt1 were examined by taking a cut through a
scan of the plate that appeared
in the original paper. We then applied a rough wavelength calibration.
Obviously, some fidelity is lost in this process but the spectra are
still useful qualitatively.     

SwSt~1 was discovered to be an emission line star by James E. Keeler in 1895
(Fleming 1895). He reported Balmer lines in emission and
compared the spectrum to that of $\eta$ Car. Although we could not find
his spectrum 
in the literature, we examined a scan from another plate
taken in 1896 which shows only 3 emission lines (H$\beta$
and [O~{\sc iii}] $\lambda$$\lambda$4959,5007) superimposed on a stellar 
continuum.
On the basis of this spectrum, SwSt~1 is referred to as P-Cygni class by 
Fleming in her log-book.
Cannon (1916), referring to these or similar spectra, also 
assigned
SwSt~1 to the P-Cygni class. These early spectra of SwSt~1 are very low
dispersion prism spectra. Although P-Cygni profiles had already
been recognized in other stars, it is doubtful that they could be
resolved on these plates. It is more
likely that SwSt~1 was classified as P-Cygni on the basis of bright Balmer 
lines
on top of a stellar continuum. Cannon (1921) mentions
the presence of [O~{\sc iii}] $\lambda$$\lambda$ 4959,5007 emission in SwSt~1. 
She notes that these lines are associated with gaseous nebulae (even
though they weren't
identified until 1928). 

Subsequently, SwSt~1 was observed spectroscopically on six distinct 
occasions:
by Payne Gaposchkin (1938, source spectra unknown); in 1940 by Swings \&
Struve (1940;
in the range 3600--6680 \AA\ at dispersions of 100 and 50 \AA /mm); in 
1942 by Swings \& Struve (1943; 3600-5300 \AA); in 1962 by Carlson and
Henize
(1979; in the range 3700--6600 \AA\ at a dispersion of 177 \AA /mm 
blue--ward of 5000 \AA\ and
94 \AA /mm red--ward of 5000 \AA ); by Aller (1977, although the date of
the 
spectrum 
is not indicated; the range of the observations is 5000-8500~\AA); in 1976
by Flower et al. (1984; in the range 4260-6710~\AA\ at a resolution of
7~\AA ); in 
1985 by de Freitas Pacheco and Feliz (1987; in
the range 3600--6930 \AA\ at resolutions of 6.5 and 0.5 \AA ); in by Leuenhagen \& Hamann (1998;
in the range 3800--5000~\AA\ and 5800-6700~\AA\ at a resolution of 1~\AA ) and 
by us in 1993 (in the range 3700--5400 \AA\ at a resolution of 0.15 \AA\ and
in the range 5400--9200~\AA\ at a resolution of 0.35~\AA ).   
                                 
Two trends since 1940 have been suggested in the literature: a
weakening of the stellar wind lines and
an increasing degree of ionization of both the stellar wind and the PN. 
The former trend was pointed out by
Carlson and Henize (1979), who noted the possible disappearance of 
several weak
stellar O~{\sc iii} and C~{\sc iii} lines between 1940 and 1962, along 
with the 
appearance of [Fe~{\sc ii}] and [Fe~{\sc iii}] lines. 
They also reported the diminution of P--Cygni profiles, from a total of 
10 lines
(as counted by Swings \& Struve, (1940)) to 4 lines, finally concluding that the 
wind's
density and mass-loss rate must have decreased between the two observations.

We looked at the plate reproductions in the original 1940 and 1962
papers and we also took a cut along the spectral direction of scanned-in
versions. To within the precision allowed by our method, we notice
no obvious change. 
It is possible that weak P-Cygni profiles were apparent to Swings \& 
Struve in the original plate material and that these features had become 
weaker by 1962 when observed by Carlson \& Henize. One might in fact expect
the star to have gradually lost its hydrogen during its early post-AGB
phase.
We have no reason to doubt the claim by Swings and Struve that H$\gamma$ and
H$\epsilon$ had P-Cygni profiles but not H$\alpha$, H$\beta$
or H$\delta$. However, we would like to point out that H$\alpha$
is generally the first line to develop an emission component and a
P-Cygni profile in the presence of a wind.
Looking at the 1940 spectrum we do notice however that the
stellar continuum is extremely irregular, while in 1960 it is much smoother.
Such irregularity, combined with a variety of emission lines could have given the
impression of troughs, which could in turn have been interpreted as P-Cygni.

The claim of Carlson \& Henize (1979) that 
[Fe~{\sc ii}] and [Fe~{\sc iii}] emission appeared sometime between 1940 and 
1962 is also somewhat suspect, since [Fe~{\sc ii}] 
emission was already reported previous to 1940, by
Payne Gaposchkin in 1938. However by far the most misleading statement about
variability reported in the literature is the claim of Carlson and Henize (1979)
that the C~{\sc iii}
$\lambda$5696 and C~{\sc iv} $\lambda$$\lambda$5801,12 lines 
are absent from the 1962 spectrum of SwSt~1.
These lines are prominent in the spectrum of 
Swings \& Struve (1942) and
also in our own spectrum. The Carlson \& Henize (1979) 1962 spectrum longward of 5000 \AA\ comes 
from a very weakly exposed plate compared to their other material.
On that plate
only the strong nebular [N~{\sc ii}] $\lambda$6583 line and H$\alpha$ 
are visible, with
He~{\sc i} $\lambda$5876 described as being barely visible.
The  C~{\sc iii} $\lambda$5696 and C~{\sc iv} $\lambda$$\lambda$5801,12 lines are 
several times 
weaker than $\lambda$5876 in our spectrum and would be well below their plate limit.
On this basis, the claim that C~{\sc iii} $\lambda$5696 and C~{\sc iv} 
$\lambda$$\lambda$5801,12 are not present in 1962 can be dismissed.

The only spectral change which seems real, comes from 
the spectroscopy
of Aller (1977). From his Figure~4 we see that C~{\sc iv} $\lambda$5470 
is much stronger and broader than the
neighbouring He~{\sc ii}/[Fe~{\sc iii}] blend at 5411~\AA . 
In the spectrum presented here, 
the strength of the two lines is comparable. However the 
C~{\sc iv} lines at 5801 and 5812~\AA\ are not stronger in the Aller
spectrum, so that no consistent trend is noticed.

de~Freitas Pacheco \& Veliz (1987) advocate a second trend, increasing ionization of both the 
stellar wind and the PN. They were not able to see any nebular
[Fe~{\sc ii}] lines in their spectra
despite the far superior resolution of their data to earlier spectra, and noticed instead 
weak lines which they attributed
to [Fe~{\sc iv}]. 
Although de Freitas Pacheco and Veliz do not show any of the spectral 
regions containing
[Fe~{\sc ii}] lines, from the quality of their published spectroscopy
we suggest that
they would not have been able to observe the weak [Fe~{\sc ii}] lines
detectable in our 1993 spectrum. 
However,
since the 1938 and 1962 spectroscopy is likely to have been of poorer 
S/N  than 
their 1985 data,
we conclude that [Fe~{\sc ii}] lines have indeed weakened
and that the fact that we do observe them in the 1993 spectrum is only because of
our higher resolution.
Based on the weakening of the nebular [Fe~{\sc ii}] lines,
we conclude that the PN {\it might} have increased its
level of ionization, although we attach a large uncertainty to this 
statement. However,
the argument that the stellar wind has increased in ionization is based
on the erroneous claim by Carlson and Henize (1979) of the absence of
the C~{\sc iv} lines at 5801 and  5812 \AA\ in 1962.

In conclusion the arguments for a significant evolution of the spectrum of SwSt~1 in the
last century are weak. The PN might have increased its level of ionization 
slightly, and there might have been some variability in the stellar wind
spectrum, but nothing major has happened to this central star in the
last 60 years and possibly in the 110 years since it was first observed.

\section{Discussion}

From the properties derived for SwSt~1 we can state the following.
Although the possible distance range of 2--4~kpc is large, even
at a distance of 2~kpc 
the stellar luminosity implied by
the models points to a central star mass at the high end of the PN
distribution.  As we
have seen, the distance issue poses a significant obstacle and does not
allow us to determine the the stellar and nebular parameters with
sufficient accuracy to establish relationships between different objects.
Better parallaxes, or expansion distances are the only way to get around
the problem. Another way would be to find [WCL] stars in the Magellanic
Clouds and use those to characterize the sample.  A wide field survey
would be needed, sensitive enough to detect emission line stars with
equivalent widths of the order of a few \AA ngstroms and for 8 m-class
telescope follow up spectroscopy. 

The stellar analysis indicates a hotter star than previously implied
(Leuenhagen \& Hamann 1998), and more in conformity with the [WC9]
spectral classification. The weakness of SwSt~1's stellar lines can be
attributed to a relatively low mass-loss rate, although the derived
value of 1.90 $\times$ 10$^{-7}$~M$_\odot$~yr$^{-1}$ should be quoted with
care because of its distance dependence. Also noteworthy is the
stellar wind oxygen abundance. This determination is reliant on a
single line of O~{\sc iii}, but no other lines exist in the available
spectral range which contradict this determination. The new
stellar evolutionary calculations of Herwig (2000) show that the oxygen
abundance of models which account for convective overshoot (which begin to
account for the severe hydrogen deficiency of WR central stars) should
have a 10-20\% oxygen mass fraction. This is in agreement with our oxygen
determination. Additionally the C/O mass ratio is in the range 2-4 
(depending on what modelling effort we rely on), showing third
dredge-up enrichment. This is not dissimilar from the values obtained for the
[WC10] stars CPD--56$^{\rm o}$8032 and He~2-113 of 3.7 and 4.1, respectively.

The PN of SwSt~1 is compact and of high density. It is likely to be stratified as
indicated by the HST H$\beta$ image being larger than the [O~{\sc iii}] image
and by
the lines from higher ionization stages being narrower.
The nitrogen abundance is not particularly high, which is surprising if
the central star is as massive as is implied by other parameters. 
Its other nebular abundances are not particularly unusual when compared
to the mean values obtained by Kingsburgh and Barlow (1994) for non-Type I
PN. We confirm again that the C/O ratio is smaller than unity (0.7 by mass), 
indicating an O-rich PN. 
This is interesting in view of the stellar C/O ratio
of 2-4, indicating active dredge up between when the nebular material was ejected the 
present time. 
A study of the C/O variation as a function of distance from the star
is underway for SwSt~1, CPD--56$^{\rm o}$5682 and He~2-113 (De Marco et al. in preparation).

The PN neutral envelope presents a
fair amount of atomic (e.g. sodium; Dinerstein et al. 1995) and
only weak molecular H$_2$ emission is observed in our SOFI spectrum.
We find that modelling this PN with a simple shell geometry and a
smooth electron density
fails no matter what stellar flux distribution is used. For a distance
of 2~kpc, our best fitting
model has a luminosity of 8900~L$_{\odot}$,
significantly higher than the re-radiated IR dust luminosity
of 1450~L$_{\odot}$, implying that either the dust shell does not
absorb and re-radiate the stellar luminosity, or else that our
stellar model is not valid. On this line we also note that the relatively low
N/H PN abundance ratio does not point to a particularly massive central star.
Additionally, if wind strength correlates with luminosity, as is the case for 
radiationally driven winds, the weakness of the wind
would also tend to point to a lower mass. 

After carefully assessing the spectral variability reported in
the literature by going back to the original publications and spectral
plates, we cannot fully corroborate any of the changes reported in
the spectrum of SwSt~1. From this
we conclude that the star has not changed significantly over the past 100
years. Certainly we have not caught SwSt~1 before the pulse that
transformed it into a hydrogen deficient star. 
The dynamical
time-scale of 150--330~yr refers to the edge of the ionized PN, although
for such a high density object, we might expect this to be only a lower
limit, since the PN material might not be fully ionized.

\section*{Acknowledgments}
OD acknowledges support through PPARC grant PPA/G/S/ 1997/00780 while at
University College London. GC and OD were supported by NASA ATP grant
NAG5/9203. OD gratefully acknowledges the financial support of Janet
Jeppson Asimov. We thank Arlo Landolt for obtaining optical
photometry of SwSt~1, while Martin Cohen and Falk Herwig
is thanked for many useful
discussions. We are also grateful to John Hillier for making his modelling code, {\sc cmfgen}, 
available to us and for the useful comments he made as the paper's referee.
This work is based, in part, on observations collected at
the European Southern Observatory, La Silla, Chile (Proposal No.
63.H--0683).

\appendix
\section{The neutral envelope of SwSt~1}
\label{app:the_neutral_envelope_of_swst1}

The Na~{\sc i} D lines in the spectrum of SwSt~1 shown in
Fig.~1 indicate a
complex neutral
envelope structure. Below we study this complex line system and
compare our findings to those of Dinerstein et al. (1995).

The Na~{\sc i} D line negative LSR radial velocity troughs in the spectrum of
SwSt~1 are unlikely to be of interstellar origin since the Galactic
radial velocity curve in the direction of SwSt~1 is positive (Fig.~1c). It
is
therefore likely that these components arise from a neutral envelope
enclosing the ionized PN.
As such they can be a good tracer of velocity components.
We identified two components per line, with radial velocities at
(--78 $\pm$ 5)~km~s$^{-1}$ and (--38 $\pm$ 2)~km~s$^{-1}$ (Fig.~1b). 
If corrected for the heliocentric radial velocity of the nebula, the
Na~{\sc i} D line emission components 
lie at (0 $\pm$ 4)~km~s$^{-1}$,
while the two troughs are measured to be at (--70 $\pm$ 5) and (--28 $\pm$ 2)~km~s$^{-1}$.
The nebular expansion velocity determined in Section~\ref{ssec:nebular_radial_and_expansion_velocities}
is (20 $\pm$ 1)~km~s$^{-1}$. Hence we would
expect absorption components of the Na D lines associated with the ionized nebula to
be found at a heliocentric radial velocity of about --20 km~s$^{-1}$.
The trough associated with the first component is at about
--28 km~s$^{-1}$, 8 km~s$^{-1}$ more negative than expected, although this discrepancy
can be almost justified by the relative uncertainties. The component at --70
km~s$^{-1}$ then has to be explained by assuming that SwSt~1's PN has a
neutral velocity component
moving at a higher velocity than the ionised 
`shell' measured by the Balmer lines.
For the [WC10] stars CPD--56$^{\rm o}$8032 
and He~2--113 (DBS97) there is a correspondence between the high velocity components of the
Na~{\sc i} D line troughs and the shape
of the emission lines of [O~{\sc i}]. The same comparison in SwSt~1 reveals that fitting
the [O~{\sc i}] lines with two Gaussians (of which one is forced to have a FWHM=40 km~s$^{-1}$)
cannot encompass components
with FWHM larger than 60 km~s$^{-1}$. This could explain a Na~{\sc i} 
D line component at about --30 km~s$^{-1}$
(like the observed component at --28 km~s$^{-1}$)
but not the trough observed at --70 km~s$^{-1}$. In conclusion,
the absorbing sodium must be associated with a different part of the
nebula from the
collisionally excited neutral oxygen.

Dinerstein et al. (1995) also found
evidence for circumstellar sodium around SwSt~1.
They measured the emission component of the Na D lines to be at --7 km~s$^{-1}$, while they determined
the absorption component to be at --27~km~s$^{-1}$ (where their
measurements were corrected for
a heliocentric nebular radial velocity of --20 km~s$^{-1}$).
These correspond within the uncertainties to our emission component at 0 km~s$^{-1}$
and absorption component at
--28 km~s$^{-1}$. They did not find a double absorption as we do, and indeed their Fig.~2 is
significantly different from our Fig.~\ref{fig:na_lines} (a),
although their claimed resolution (0.20 \AA ) is similar to our own.
It is not possible at present to resolve this discrepancy.
However they did mention the possibility of the existence of higher
velocity components in the
PN envelope and they compared this effect to the fast low ionization
outflows (FLIERs) described by
Balick et al. (1993) and Balick (1994).

\section{The optical and near-IR spectra of SwSt~1}
\label{app:the_optical_and_near_ir_spectra_of_swst1}

\begin{figure*}
\vspace{22cm}
\includegraphics{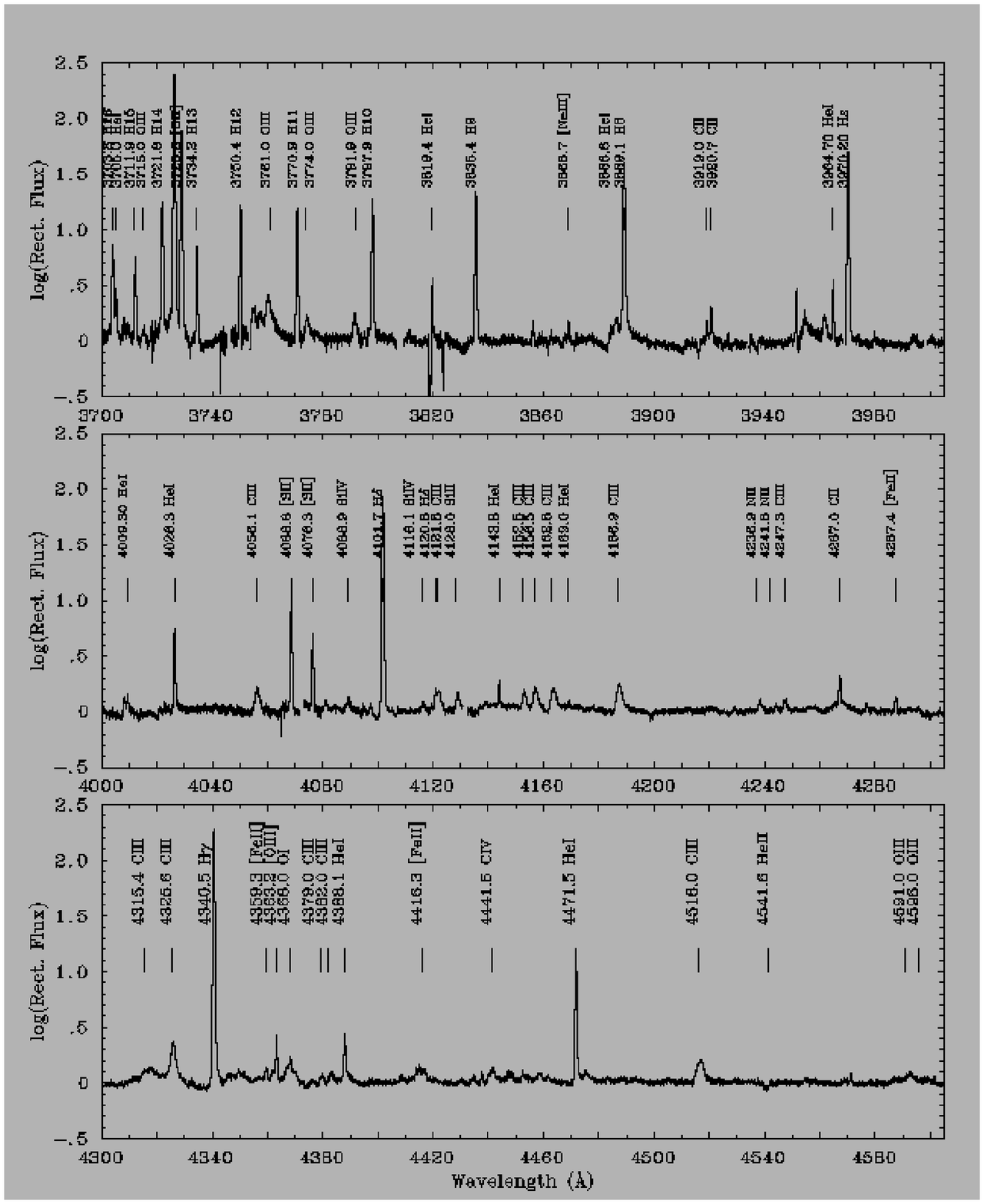}
\caption
{The AAT spectrum of SwSt~1 between 3700 and 4605 \AA .}
\label{swst_atl_1}
\end{figure*}

\begin{figure*}
\vspace{22cm}
\includegraphics{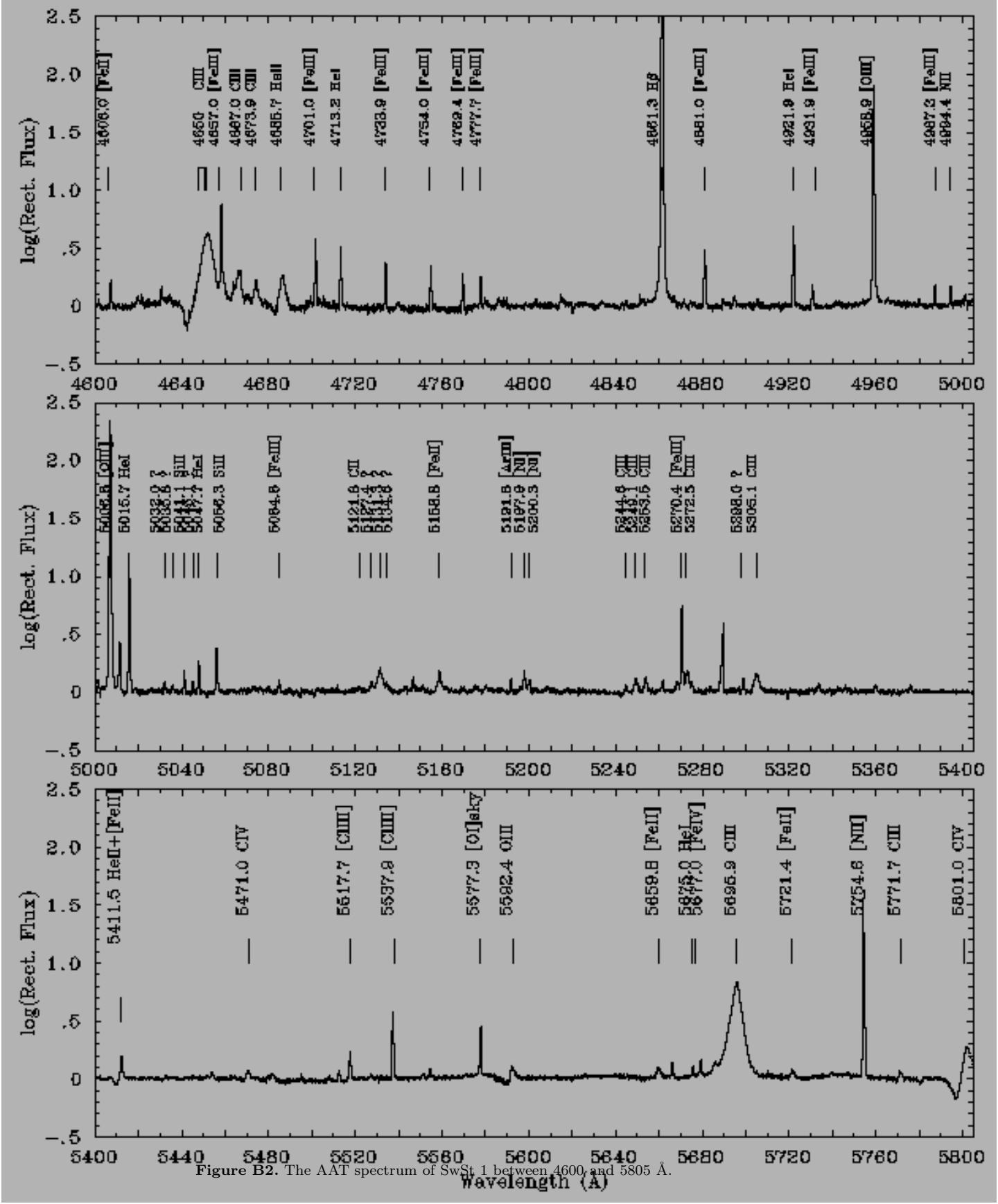}
\caption{The AAT spectrum of SwSt~1 between 4600 and 5805 \AA .}
\label{swst_atl_2}
\end{figure*}

In this Appendix we present the UCL Echelle spectrum of SwSt~1 obtained at the
Anglo--Australian Telescope in May 1993
(Figs.~\ref{swst_atl_1}-\ref{swst_atl_5})
as well as our near-IR NTT SOFI spectrum (Fig.~\ref{fig:SOFI}; details of
the data
acquisition and reduction are presented in
Section~\ref{sec:observations}).
Due to the strength of the nebular lines compared to the stellar lines, 
we present the rectified optical spectrum on a log scale.

In Table~\ref{tab:atlas}
we present a list of the lines we identified in the optical spectrum of SwSt~1. 
In Column~4 we indicate
the origin of the line ({\it s} for stellar, {\it n} for nebular, {\it is} 
for interstellar), while in Column~5
we list the initials of the authors who have identified those lines in the past.
A discussion of the stellar and nebular components of SwSt~1's optical
spectrum
has been presented in Secs.~\ref{ssec:the_stellar_spectrum} and
\ref{ssec:nebular_line_fluxes}.

The near-IR SOFI spectrum is dominated by strong nebular emission lines of
H\,{\sc i} (e.g. P$\beta$, Br$\gamma$) and He\,{\sc i} (1.083$\mu$m,
2.058$\mu$m), plus [S\,{\sc iii}] 9530\AA. As reported in
Sec.~\ref{ssec:near_ir_spectroscopy}, we have detected
$H_{2}$ (at 2.1218$\mu$m ($v$=1--0, S(1)) for the first time from this 
young PN.

\begin{table}
\caption{Lines identified in the spectrum of SwSt~1 between 3700 and 9700 \AA .}
\begin{tabular}{lllll}
Ion & Wavelength & Multiplet & Origin &Comments \\
    & (\AA )     &  No.      & &\\
H16               &3703.8 &              &n&SS,CH               \\
He~{\sc i}        &3705.0 &  25   &n& CH               \\
H15               &3711.9 &              &n&SS,CH               \\
O~{\sc iii}       &3715.0 &  14   &s &SS               \\
H14               &3721.8 &              &n&SS,CH,dFV            \\
$[$O~{\sc ii}]    &3726.2 & 1F  &n&SS,CH,dFV            \\
$[$O~{\sc ii}]    &3728.9 & 1F  &n&SS,CH,dFV            \\
H13               &3734.2 &              &n&SS,CH               \\
H12               &3750.4 &              &n&SS,CH               \\
O~{\sc iii}       &3761.0 &  30   &s &SS               \\
H11               &3770.9 &              &n&SS,CH,dFV            \\
O~{\sc iii}       &3774.0 &  2    &s &                 \\
O~{\sc iii}       &3791.3 &  2    &s &                 \\
O~{\sc iii}       &3791.9 &  2    &s &SS               \\
H10               &3797.9 &              &n&SS,CH,dFV            \\
He~{\sc i}        &3819.4 &  22   &n  &SS,CH               \\
H9                &3835.4 &       &n&SS,CH,dFV            \\     
$[$Ne~{\sc iii}]    &3967.7 &       &n&\\
He~{\sc i}        &3888.7 & 2   &n        &SS\\
H8                &3889.1 &              &n&SS,CH,dFV            \\
C~{\sc ii}        &3919.0 & 4   &n&CH\\
C~{\sc ii}        &3920.7 & 4   &n&CH\\
He~{\sc i}        &3964.7 &  5    &n   &SS,CH               \\
H$\epsilon$       &3970.2 &          &n&SS,CH,dFV            \\
He~{\sc i}        &4009.4 & 55    &n&\\
He~{\sc i}        &4026.3 &  18   &n  &SS,CH               \\
C~{\sc iii}       &4056.1 &  24   &s &SS,CH               \\
$[$S~{\sc ii}]    &4068.6 & 1F  &n   &SS,CH,dFV            \\
$[$S~{\sc ii}]    &4076.3 & 1F  &n  &SS,CH,dFV            \\
H$\delta$         &4101.7 &       &n&SS,CH,dFV            \\
He~{\sc i}        &4120.8 & 16 &n\\
C~{\sc iii}       &4121.8 &  17   &s &                 \\
Si~{\sc ii}       &4128.0 &  3     &s &  CH               \\
Si~{\sc ii}       &4130.9 & 3 &s\\
N~{\sc ii}        &4136.9 &   & s/n?&\\
He~{\sc i}        &4143.8 &  53   &n&SS,CH               \\
C~{\sc iii}       &4152.5 &  21   &s &                 \\
C~{\sc iii}       &4156.5 &  21   &s &SS               \\
C~{\sc iii}       &4162.8 &  21   &s &                 \\
He~{\sc i}        &4169.0 &  52   &n&                 \\
C~{\sc iii}       &4186.9 &  18   &s &SS,CH               \\
N~{\sc ii}        &4241.8 &       &s/n?&\\
C~{\sc iii}       &4247.3 &  11   &s &                 \\
C~{\sc ii}        &4267.0 &  6    &n+s&CH,dFV               \\
$[$Fe~{\sc ii}]   &4287.4 &7F     &n&  CH               \\
C~{\sc iii}       &4315.4 &  48   &s &                 \\
C~{\sc iii}       &4325.6 &  7    &s & CH                \\
H$\gamma$         &4340.5 &            &n&SS,CH,dFV            \\
$[$Fe~{\sc ii}]   &4359.3 &7F     &n     &  CH               \\
$[$O~{\sc iii}]   &4363.2 &2F     &n    &SS,CH,dFV            \\
O~{\sc i}         &4368.0 &       &n &dFV               \\   
C~{\sc iii}       &4379.0 &  14   &s &                 \\
C~{\sc iii}       &4382.0 &  14   &s&                 \\
He~{\sc i}        &4388.1 &  51   &n &SS,dFV,CH            \\
$[$Fe~{\sc ii}]   &4416.3 &  6F   &n  &   CH              \\
\end{tabular}
\end{table}

\begin{table}
\begin{tabular}{lllll}
Ion & Wavelength & Multiplet & Origin & Comments \\
    & (\AA )     & No.       &        &\\
C~{\sc iv}        &4441.5 &       &s &                    \\
C~{\sc iii}       &4650.3 &  1    &s &SS,CH               \\
C~{\sc iii}       &4651.5 &  1    &s &SS,CH               \\
$[$Fe~{\sc iii}]  &4657.0 &  3F   &n  &dFV,CH               \\
C~{\sc iii}       &4663.6 &  5    &s&\\
C~{\sc iii}       &4667.0 &  5    &s &SS               \\
C~{\sc iii}       &4673.9 &  5    &s &                 \\
He~{\sc ii}       &4685.7 &  3-4  &s  &SS,CH               \\
$[$Fe~{\sc iii}]  &4701.0 &  3F   &n    &CH,dFV               \\
He~{\sc i}        &4713.2 &  12   &n&   CH,dFV              \\
$[$Fe~{\sc iii}]  &4733.9 &  3F   &n &CH,dFV               \\
$[$Fe~{\sc iii}]  &4754.0 &  3F     &n&dFV               \\
$[$Fe~{\sc iii}]  &4769.4 &  3F   &n&\\
$[$Fe~{\sc iii}]  &4777.7 &  3F   &n&\\
H$\beta$          &4861.3 &       &n   &SS,CH,dFV            \\
$[$Fe~{\sc iii}]  &4881.0 &  2F   &n&\\
He~{\sc i}        &4921.9 &  48   &n &CH,dFV               \\
$[$Fe~{\sc iii}]  &4930.5 &  1F   &n&\\
$[$O~{\sc iii}]   &4958.9 &  1F   &n  &SS,CH,dFV            \\
$[$Fe~{\sc iii}]  &4987.3 &       &n&\\
N~{\sc ii}]      &4994.4 &       &s/n&\\
$[$O~{\sc iii}]   &5006.8 &  1F   &n &SS,CH,dFV            \\
He~{\sc i}        &5015.7 &  4    &n&dFV               \\
?                 &5032.0 &       & &\\
?                 &5035.8 &       & &\\
?                 &5041.1 &&&\\
He~{\sc i}        &5047.7 &  47   &n&                 \\
Si~{\sc ii}       &5056.0 &   5   &n?&\\
$[$Fe~{\sc iii}]  &5084.8 &&&\\
C~{\sc ii}        &5121.8 &&&\\
?                 &5130.8 &       &s& \\
$[$Fe~{\sc ii}]   &5158.8 &  19F  &n&\\
$[$Ar~{\sc iii}]  &5191.8 &  3F   &n &                 \\        
$[$N~{\sc i}]     &5197.9 &  1F   &n &                 \\
$[$N~{\sc i}]     &5200.3 &  1F   &n &                 \\
C~{\sc iii}       &5244.6 &  4    &s &                 \\
C~{\sc iii}       &5249.1 &  23   &s &                 \\
C~{\sc iii}       &5253.5 &  4    &s &                 \\
$[$Fe~{\sc iii}]  &5270.0 &  1F   &n  &dFV               \\
C~{\sc iii}       &5272.5 &   4   &s &                 \\
?                 &5298.0&&&\\
C~{\sc iii}       &5305.1 &   46  &s &                 \\
He~{\sc ii}       &5411.5 &   4-7 &s&                 \\
C~{\sc iv}        &5471.0 &       &s&                 \\
$[$Cl~{\sc iii}]  &5517.7 &  1F &n   &dFV               \\
$[$Cl~{\sc iii}]  &5537.9 &  1F   &n   &dFV               \\
$[$O~{\sc i}]     &5577.3 &  3F   &n&  sky               \\
OIII              &5592.4 &  5    &s& \\
$[$Fe~{\sc ii}]   &5659.8 &       &n&\\
He~{\sc i}        &5675.0 &   ?   &n&                 \\
C~{\sc iii}       &5695.9 &   12  &s &SS               \\
$[$Fe~{\sc ii}]   &5721.4 &  33F  &n &\\
$[$N~{\sc ii}]    &5754.6 &  3F &n  &SS,dFV            \\
C~{\sc iii}       &5771.7 &   ?   &s&\\
C~{\sc iv}        &5801.3 &   1   &s &SS,dFV               \\
C~{\sc iv}        &5812.0 &   1   &s&  dFV               \\
C~{\sc iii}       &5826.4 &   22  &s&                 \\
C~{\sc iii}       &5857.9 &   20  &s&\\
C~{\sc iii}       &5871.6 &   20  &s&\\
He~{\sc i}        &5875.7 &   11  &n &SS,CH,dFV        \\
Na~{\sc i} D2     &5889.9 &       &n+is &                 \\
C~{\sc iii}       &5894.2 &   20  &s    &\\
Na~{\sc i} D1     &5895.9 &       &n+is &                 \\
Si~{\sc ii}       &5957.6 &  4    &s/n?&\\
C~{\sc iii}       &6205.5 &   17  &s     &                 \\
$[$O~{\sc i}]     &6300.3 &   1F  &n     &SS,dFV            \\
\end{tabular}
\end{table}
\begin{table}
\begin{tabular}{lllll}
Ion & Wavelength & Multiplet & Origin & Comments \\
    & (\AA )     & No.       &        &\\
$[$S~{\sc iii}]   &6312.1 &   3F  &n     &SS,dFV            \\
$[$O~{\sc i}]     &6363.8 &   1F  &n     &dFV               \\            
$[$N~{\sc ii}]    &6548.0 & 1F    &n     &SS,CH,dFV            \\
H$\alpha$         &6562.8 &       &n     &SS,CH,dFV            \\
Si~{\sc ii}       &6347.1 &  2    &n/s?&\\
Si~{\sc ii}       &6371.4 &  2    &n/s?&\\
C~{\sc ii}        &6578.1 &   2   &n     &                 \\
$[$N~{\sc ii}]    &6583.4 & 1F      &n     &SS,CH,dFV            \\
He~{\sc i}        &6678.1 &         &n     &SS,dFV               \\
$[$S~{\sc ii}]    &6716.5 &   2F    &n     &dFV               \\
C~{\sc iii}       &6727.4 &   3     &s     &                \\
$[$S~{\sc ii}]  &6730.8 &    2F   &n     &dFV               \\
C~{\sc iii}     &6744.4 &   3     &s     &                 \\
C~{\sc iii}     &6762.2 &   3     &s     &                 \\
C~{\sc iii}     &7037.2 &   6.01  &s     &                 \\
He~{\sc i}      &7065.7 &   10    &n     &                 \\
$[$Ar~{\sc iii}]&7135.8 &   1F    &n     &                 \\
C~{\sc ii}      &7231.3 &   3     &n     &                 \\
C~{\sc ii}      &7236.4 &   3     &n     &                 \\
He~{\sc i}      &7281.3 &   45    &n     &                 \\
$[$O~{\sc ii}]  &7319.9 &   2F    &n     &                 \\
$[$O~{\sc ii}]  &7330.1 &   2F    &n     &                 \\
C~{\sc iii}     &7486.5 &   41    &s     &                 \\
C~{\sc iii}     &7586.4 &   2.02  &s     &                 \\
C~{\sc iii}     &7707.4 &   10.01 &s     &                 \\
$[$Ar~{\sc iii}]&7751.1 &   1F    &n     &                 \\
C~{\sc iii}&7771.8 &26&s\\
C~{\sc iii}&7780.4 &26&s\\
C~{\sc iii}&7796.0 &26&s\\
C~{\sc iii}&8196.2 &  43          &s     &         \\
H3-20      &8392.4 &              &n     &         \\
H3-19      &8413.3 &              &n     &         \\
H3-18      &8438.0 &              &n     &         \\
H3-17      &8467.3 &              &n     &         \\
$[$Cl~{\sc iii}]&8500.0 &           &n     &         \\                 
C~{\sc iii}&8500.3 &   1.01       &s     &         \\                 
H3-16      &8502.5 &              &n     &         \\
H3-15      &8545.4 &              &n     &         \\
$[$Cl~{\sc ii}]     &8578.7 &     &n     &                 \\
H3-14      &8598.4 &              &n     &         \\
$[$Fe~{\sc ii}]&8617.0&           &n     &         \\
H3-13      &8665.1 &              &n     &         \\
H3-12      &8750.5 &              &n     &         \\
H3-11      &8862.7 &              &n     &         \\
H3-10      &9014.9 &              &n     &         \\
$[$S~{\sc iii}]     &9068.9 &1F   &n     &                 \\
H3-9       &9229.0 &              &n     &         \\
$[$S~{\sc iii}]     &9531.0 &1F   &n     &                 \\
\multicolumn{5}{l}{SS=Swings and Struve 1940; range 3600--6680 \AA ; dispersion}
\\
\multicolumn{5}{l}{100 and 50 \AA /mm.}\\
\multicolumn{5}{l}{CH=Carlson and Henize 1979; observations taken in 1962,}\\
\multicolumn{5}{l}{range 3700--6600 \AA , resolution 177 \AA /mm up to} \\
\multicolumn{5}{l}{5000, 94 \AA /mm after.}\\
\multicolumn{5}{l}{dFV=de Freitas Pacheco and Veliz 1984; range 3600--5900 \AA }
\\
\multicolumn{5}{l}{resolution 6.5 \AA\ and 0.5 \AA ;NB some lines observed by th
em} \\                                  
\multicolumn{5}{l}{might be missing.}\\
\label{tab:atlas}
\end{tabular}
\end{table}

\begin{table}
\caption{Principal near-IR line fluxes, $F_{\lambda}$, 
obtained from NTT-SOFI observations of SwSt~1, together with 
de-reddened fluxes relative to I(H$\beta$)=1.50 $\times 10^{-10}$ 
erg\,cm$^{-2}\,s^{-1}$, assuming E(B--V)=0.46 mag. Uncertain
values are given in parenthesis (due to poor atmospheric transmission).}
\renewcommand\arraystretch{0.7}
\begin{small}
\begin{center}
\begin{tabular}{lllll}
$\lambda_{\rm obs}$ & Identification & $\lambda_{\rm lab}$ & $F_{\lambda}$ & 100$\times$$I_{\lambda}$ \\
($\mu$m) &     & ($\mu$m) & (erg\,cm$^{-1}\,s^{-1}$) & /I(H$\beta$)\\
0.954  &[SIII]  &0.953& 7.56$\times 10^{-11}$  & 90.7 \\
0.971  &CIII(3p-3d)&0.971&2.30$\times 10^{-12}$& 2.7  \\
1.005  &P$\delta$&1.005& 4.86$\times 10^{-12}$&  5.6\\
1.033  &        &     &  1.24$\times 10^{-12}$&  1.3\\
1.084  &HeI(2s-2p)&1.083&5.69$\times 10^{-11}$&60.8 \\
1.094  &P$\gamma$&1.094& 8.46$\times 10^{-12}$&  9.0\\
1.198  &CIII(4s-4p)&1.198&3.7$\times 10^{-13}$& 0.4 \\
1.282  &P$\beta$&1.282& 1.56$\times 10^{-11}$& 14.8\\
1.317  &        &     &  2.6$\times 10^{-13}$&  0.2\\   
1.520  &HI(4-20)&1.519&  1.4$\times 10^{-13}$&  0.1\\
1.527  &HI(4-19)&1.526&  1.5$\times 10^{-13}$&  0.1\\
1.535  &HI(4-18)&1.534&  2.3$\times 10^{-13}$&  0.2\\
1.544  &HI(4-17)&1.544&  2.2$\times 10^{-13}$&  0.2 \\
1.556  &HI(4-16)&1.556&  2.5$\times 10^{-13}$&  0.2\\
1.571  &HI(4-15)&1.570&  3.3$\times 10^{-13}$&  0.3\\
1.589  &HI(4-14)&1.588&  4.1$\times 10^{-13}$&  0.3\\
1.612  &HI(4-13)&1.611&  4.4$\times 10^{-13}$&  0.4\\
1.641  &HI(4-12)&1.641&  6.5$\times 10^{-13}$&  0.5\\
1.681  &HI(4-11)&1.681&  8.0$\times 10^{-13}$&  0.7\\
1.701  &HeI(3p-4d)&1.700&1.5$\times 10^{-13}$&  0.1\\
1.737  &Br$\zeta$&1.736& 9.1$\times 10^{-13}$&  0.7\\
1.818  &Br$\epsilon$&1.817&3.5$\times 10^{-12}$& 2.8 \\
1.875  &P$\alpha$&1.875&(1.78$\times 10^{-11}$)& (14.0)\\
1.945  &Br$\delta$&1.945&1.9$\times 10^{-12}$&  1.5\\
2.058  &HeI(2s-2p)&2.058&1.4$\times 10^{-12}$&   1.0\\
2.122  &H$_2$(0-1, S(1))&2.122 & 1.5$\times 10^{-13}$&  0.1 \\
2.166  &Br$\gamma$&2.166&3.2$\times 10^{-12}$&  2.4 \\
\label{table:nearir_fluxes}
\end{tabular}
\end{center}
\end{small}
\end{table}

\begin{figure*}
\vspace{22cm}
\includegraphics{gks74.ps.2-new}
\caption{The AAT spectrum of SwSt~1 between 5800 and 7005 \AA .}
\label{swst_atl_3}
\end{figure*}

\begin{figure*}
\vspace{22cm}
\includegraphics{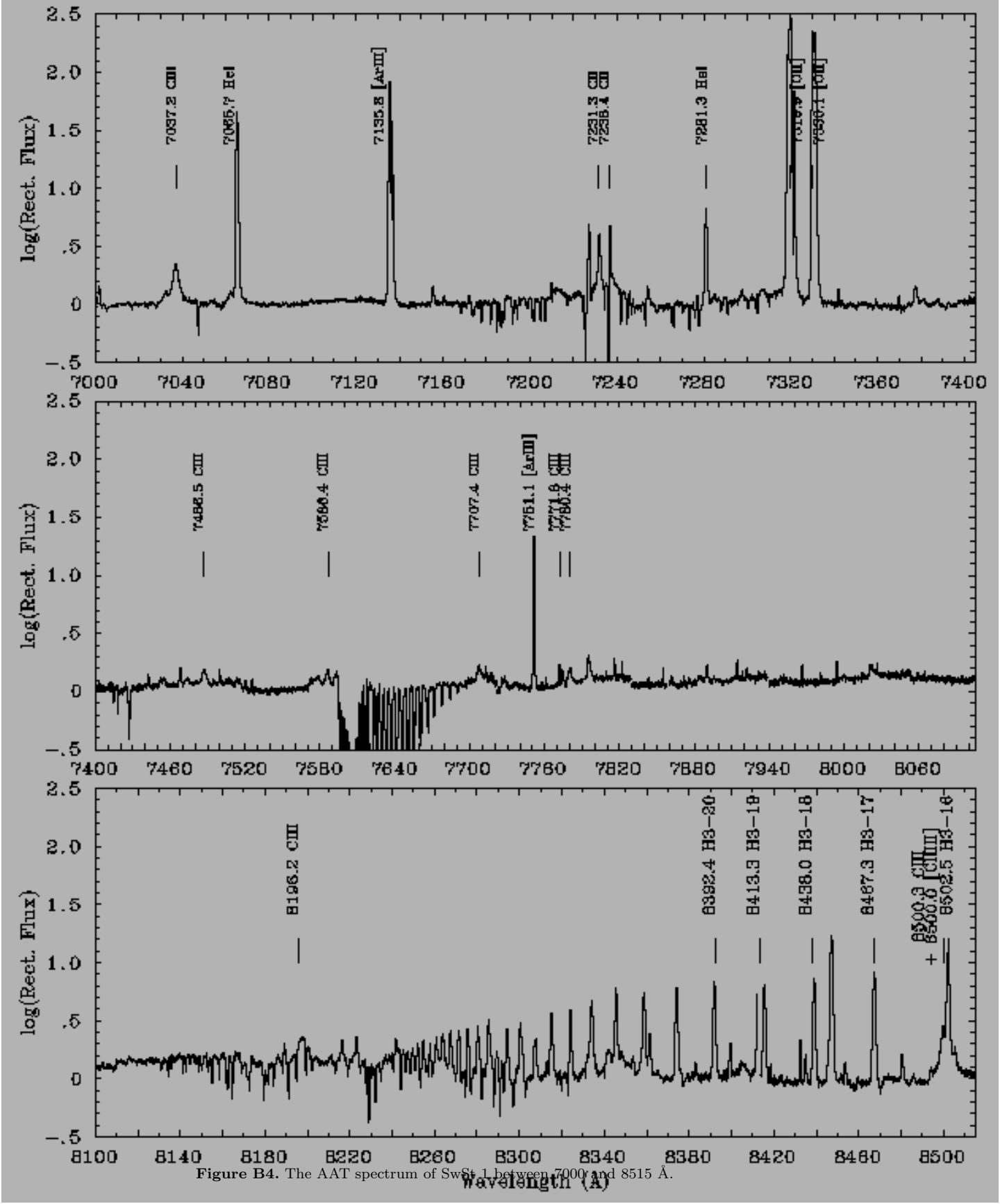}
\caption{The AAT spectrum of SwSt~1 between 7000 and 8515 \AA .}
\label{swst_atl_4}
\end{figure*}

\begin{figure*}
\vspace{22cm}
\includegraphics{gks74.ps.4-new}
\caption{The AAT spectrum of SwSt~1 between 8500 and 9700 \AA .}
\label{swst_atl_5}
\end{figure*}             

\begin{figure*}
\vspace{21cm}
\includegraphics{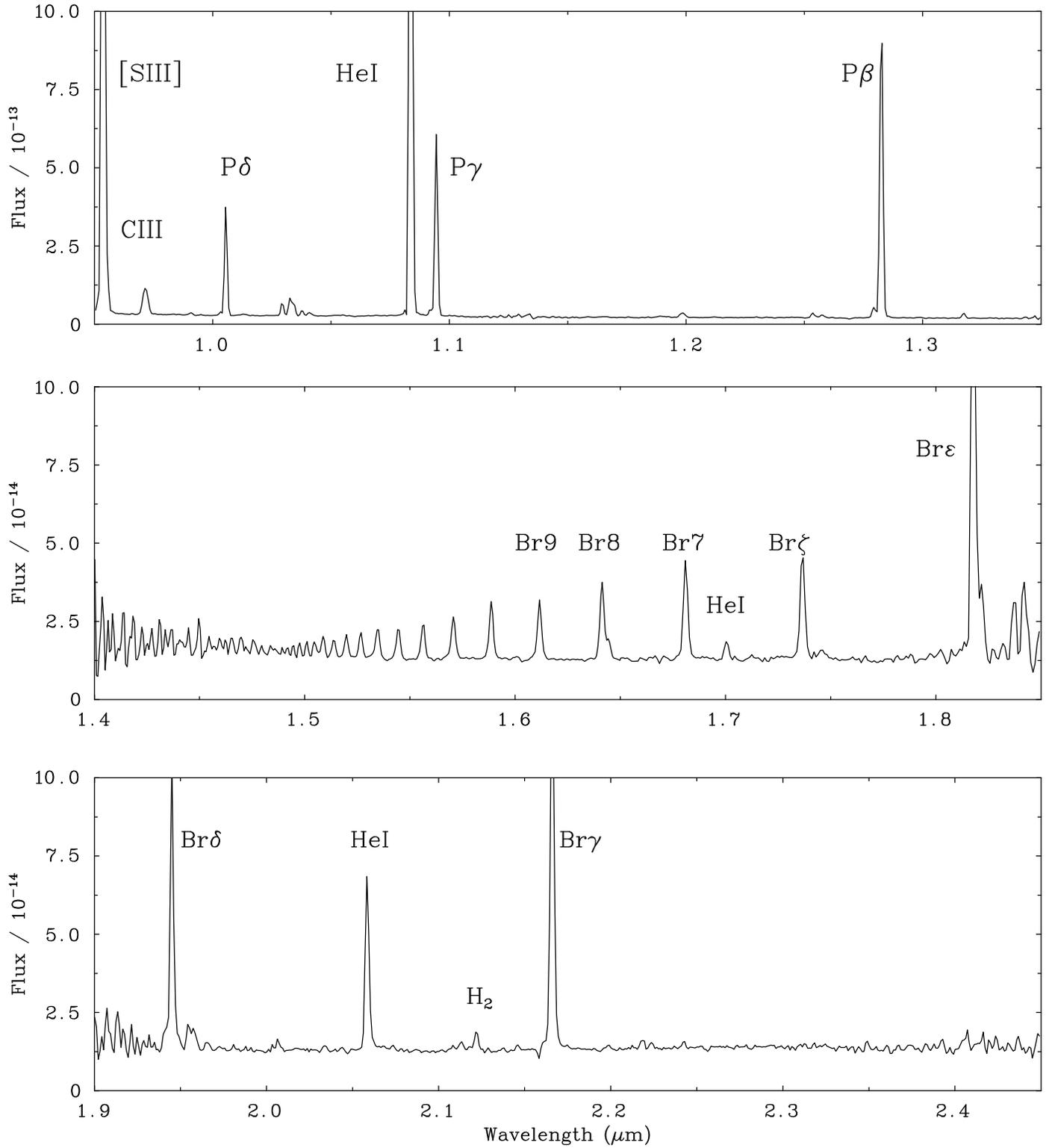}
\caption{The SOFI spectrum of SwSt~1.}
\label{fig:SOFI}
\end{figure*}

\bsp

\label{lastpage}

\end{document}